\documentclass[useAMS,usenatbib]{mn2e}
\usepackage{graphicx,natbib}
\usepackage{amssymb}

\def\ms{\mbox{$M_\odot$}}

\title[Sequential star formation in G174+2.5?]{A possible sequential star formation in the giant molecular cloud G174+2.5}

\author[D. Camargo, C. Bonatto and E. Bica]{D. Camargo$^1$, C. Bonatto$^1$ and E. Bica$^1$\\
$^1$ Departamento de Astronomia, Universidade Federal do Rio Grande do Sul, 
Av. Bento Gon\c{c}alves 9500\\
Porto Alegre 91501-970, RS, Brazil}

\begin{document}

\pagerange{\pageref{firstpage}--\pageref{lastpage}}

\maketitle

\label{firstpage}

\begin{abstract}
We investigate the nature of 14 embedded clusters (ECs) related to a group of four H II regions Sh2-235, Sh2-233, Sh2-232, and Sh2-231 in the giant molecular cloud G$174+2.5$. Projected towards the Galactic anticentre, these objects are a possible example of the \textit{collect and collapse} scenario. We derive astrophysical parameters (\textit{age, reddening, distance, core and cluster radii}) for the ECs and investigate the relationship among their parameters.
Parameters are derived with field decontaminated 2MASS colour-magnitude diagrams (CMDs) and stellar radial density profiles (RDPs).
The CMDs of these young clusters are characterised by a poorly-populated main sequence and a significant number of pre-main sequence stars, affected by differential reddening. The ECs are KKC 11, FSR 784, Sh2-235 E2, Sh2-235 Cluster, Sh2-233SE Cluster, BDSB 73, Sh2-235B Cluster, BDSB 72, BDSB 71, Sh2-232 IR, PCS 2, and the newly found clusters CBB 1 and CBB 2. We were able to derive fundamental parameters for all ECs in the sample. Structural parameters are derived for FSR 784, Sh2-235 Cluster and Sh2-235E2.
\end{abstract}

\begin{keywords}
({\it Galaxy}:) open clusters and associations:general; {\it Galaxy}: stellar content, \textit{collect and collapse}, and sequential star formation; {\it Galaxy}: structure
\end{keywords}

\section{Introduction}
\label{Intro}
Most stars form after the gravitational collapse of massive and dense gas clumps inside  giant molecular clouds (GMCs), with collapsing clumps forming embedded clusters (ECs). However, supernova explosions, H II region expansion due to massive-stars, UV radiation, and stellar winds, may disrupt GMCs completely, on a timescale of a few $10^7$ yr \citep[][]{Elmegreen00,Bonnell06}.  In this context, \citet{Hartmann01} point out that stars older than $\approx5$ Myr are not found associated with molecular gas, and \citet{Allen07} suggest that the primordial gas of ECs disperses in $3\,-\,5$ Myr \citep[][]{Leisawitz89, Proszkow09}. For \citet{Lada03}, the duration of the embedded phase is $2\,-\,3$ Myr. On the other hand, the parent GMCs are very disruptive environments for ECs. \citet{Lada03} estimate that only 4 - 7\% of them survive for more than 40 Myr (\textit{infant mortality}), but the appearance of bound and unbound clusters are indistinguishable for clusters younger than 10 Myr. In other words, stars are born in star clusters embedded in GMCs but, as a consequence of the disruptive mechanisms, most end up as part of the field-star population after cluster disruption.

ECs can be partially or fully immersed in embryonic molecular clouds and H II regions. The youngster are located in gas clumps and the most evolved are often linked with H II regions or other nebulae \citep{Leisawitz89}. Often, these H II regions, excited by fast winds from massive OB stars, expand into the molecular cloud triggering sequential star formation. In this sense, the \textit{collect and collapse} model \citep{Elmegreen77, Whitworth94}  suggests that H II regions expand accumulating material between the ionisation and the shock fronts. This material, as a consequence of the shocks, becomes unstable and fragments into several cores, triggering star formation in multiple protocluster regions. Another possible process is the ``radiation-driven implosion" model, in which the expanding H II region compresses the existing molecular clumps, the density increase exceeding the critical mass and the clump collapses \citep{Lefloch94}. In both scenarios the massive stars trigger a second generation of cluster formation \citep[see also,][]{Fukuda00, Hosokawa05, Deharveng05, Dale07}. In any case, GMC observations indicate that there often occurs a multiple cluster formation as a result of winds of OB stars or expanding H II regions \citep{Allen07}. 

\begin{figure}
\begin{minipage}[b]{1.0\linewidth}
\includegraphics[width=\textwidth]{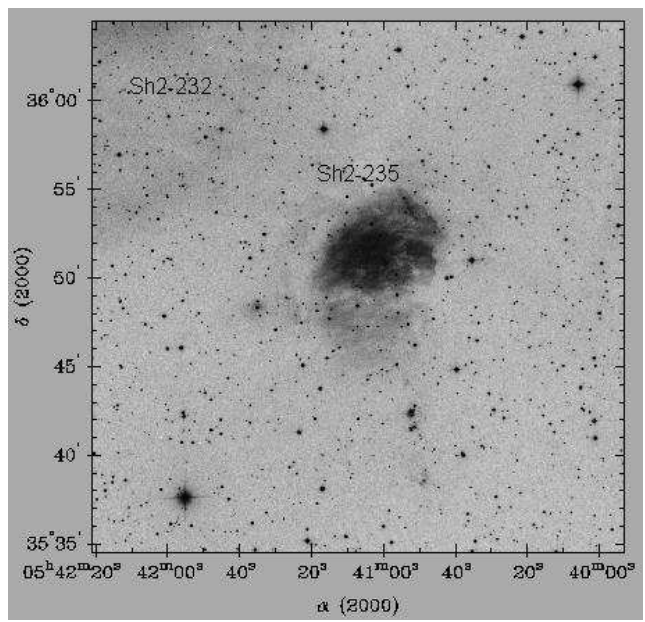}
\end{minipage}\hfill
\caption[]{XDSS R image ($30'\times30'$) of the Sh2-235 HII region. The image also shows part of the Sh2-232 HII region (top left).}
\label{fig:01}
\end{figure}

\begin{table*}
{\footnotesize
\tiny
\caption{Cross-identifications of the embedded clusters in G$174+2.5$.}
\renewcommand{\tabcolsep}{5.0mm}
\renewcommand{\arraystretch}{1.1}
\begin{tabular}{lllllllll}
\hline
\hline
Desig\#1&Desig\#2&Desig\#3&Desig\#4&Desig\#5&D (')&Ref.\\
($1$)&($2$)&($3$)&($4$)&($5$)&($6$)&($7$)\\
\hline
KKC\,11&FSR\,788&Sh2-235\,East1&-&$-$&4.5&7,\,8,\,10\\
FSR\,784&Koposov\,7&Sh2-235North-West&-&-&3.2&8,\,9,\,10\\
Sh2-235\,E2&-&-&-&-&1.3&6\\
Sh2-235\,Cl.&CSSS\,11&Sh2-235\,Central&-&-&1.4&1,\,10\,\\
BDSB\,73&KSTW\,1&&-&&0.7&5,\,8\\
Sh2-235B\,Cl.&CSSS\,10&Hodapp\,18&Sh2-235ABC&-&1.6&1,\,2,\,10\\
BDSB\,72&-&-&-&-&2.0&6\\
BDSB\,71&-&-&-&-&0.8&6\\
Sh2-232\,IR&&&-&&1.0&10\\
PCS\,2&DB2001-24&KKC\,9\,NE&-&-&1.0&3,\,4,\,7\\
Sh2-233\,SE\,cl.&Hodapp\,15&PCS\,1&KKC\,9\,SW&Sh2-233\,IR&1.1&2,\,3,\,6,\,10\\
$G173.58+2,45$\,Cl.&$IRAS\,05361+3539$\,Cl.&-&-&-&0.5&5,\,10\\
CBB\,1&-&-&-&-&0.3&11\\
CBB\,2&-&-&-&-&1.0&11\\
\hline
\end{tabular}
\begin{list}{Table Notes.}
\item Cols. $(1-5)$ show cross-identification and col. $(7)$ references for parameter determinations. The references are: 1 - \citet{Carpenter93}; 2 - \citet{Hodapp94}; 3- \citet{Porras00}; 4 - \citet{Dutra01}; 5 - \citet{Shepherd02};  6 - \citet{Bica03}; 7 - \citet{Kumar06}; 8 -\citet{Froebrich07} ; 9 -\citet{Koposov08}; 10 - \citet{Kirsanova08}; 11 - This work.
\end{list}
\label{tab2}
}
\end{table*}

Star cluster formation is the preferential mode of star formation, and ECs may be responsible for  $70\%$-$90\%$  of all stars formed in GMCs. 
However, the rapid expulsion of the primordial gas by winds of OB stars and supernova explosions disrupt most clusters very early. As a consequence of rapid gas expulsion, the stellar orbits cannot adjust to the new potential, and probably give rise to an unbound association. N-body simulations show that in this phase ($10$ - $30$ Myr) the cluster expands in all scales reaching for virialization. 
The fate of a cluster is determined by the star formation efficiency (SFE). If the gas is removed slowly the cluster will remain bound as long as the SFE is higher than $30\%$, but if the gas is removed rapidly, the SFE needs to be higher than $50\%$. As a consequence of \textit{infant mortality}, the number of optically detected clusters is significantly smaller than that of ECs \citep[][]{Lada84, Verschueren90, Lada03, Goodwin06}.

Young clusters in expanding H II regions are characterised by a significant population of low-mass pre-main sequence (PMS) stars together with variable fractions of massive OB stars \citep{Saurin10, Bonatto11}. \citet{Dobashi01} show that protostars associated with H II regions are more luminous than those in molecular clouds away from them, which indicates that H II regions favour massive stars or cluster formation in neighbouring molecular clouds. \citet{Deharveng03} suggest that clusters formed from these processes are still deeply embedded in H II regions, and are not dispersed. 

The observed colour magnitude diagram (CMD) of an EC in general presents a relatively vertical, poorly-populated main sequence (MS) and a well developed PMS with a large population of faint and red stars \citep{Bonatto09}. This feature is evident when the raw photometry and decontaminated CMDs are compared to field star extractions in the present analysis.
The low mass stars remain in the PMS phase for $\approx30$ Myr \citep{Gouliermis08}.

In this work we investigate properties of the star clusters embedded in the GMC $G174+2.5$, related to a group of H II regions located in the Perseus arm towards the Galactic anticentre. A star forming complex with numerous ECs minimises uncertainties owing to the distance in common.
This paper is organised as follows. In Sect.~\ref{sec:2} we provide general data on the target clusters. In Sect.~\ref{sec:3} we present the 2MASS photometry and introduce tools employed in the CMD analyses, especially the field star decontamination algorithm. Sect.~\ref{sec:4} is dedicated to the cluster structure. In Sect.~\ref{Mass} we estimate cluster masses. In Sect. ~\ref{sec:5} we investigate the relationship among the derived parameters. Finally, in Sect.~\ref{sec:6} we present the concluding remarks.

\begin{figure*}
\begin{minipage}[b]{1.0\linewidth}
\includegraphics[width=\textwidth]{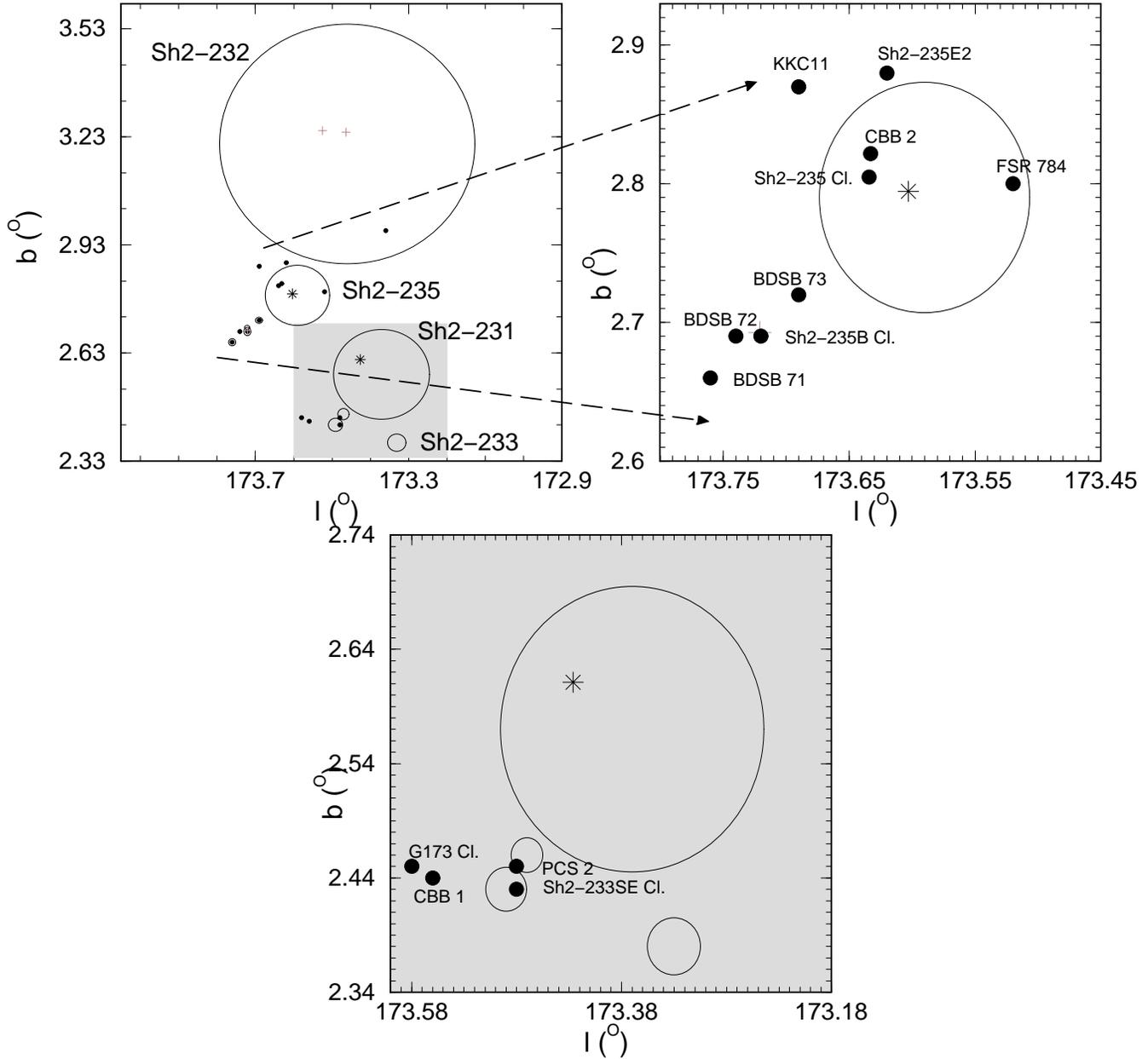}
\end{minipage}\hfill
\caption[]{Top left panel: schematic map of the nebulae (open circles). Filled circles are ECs, asterisks are O stars, and plus signs are B stars. Top right panel: zoom of the region that contains the HII region Sh2-235. Bottom panel: zoom of the gray region in the left top panel that contains the nebulae Sh2-231, Sh2-233, RNO49 and Sh2-233SE.}
\label{fig:02}
\end{figure*}

\begin{figure*}
   \centering
   \includegraphics[scale=0.32,viewport=0 0 490 460,clip]{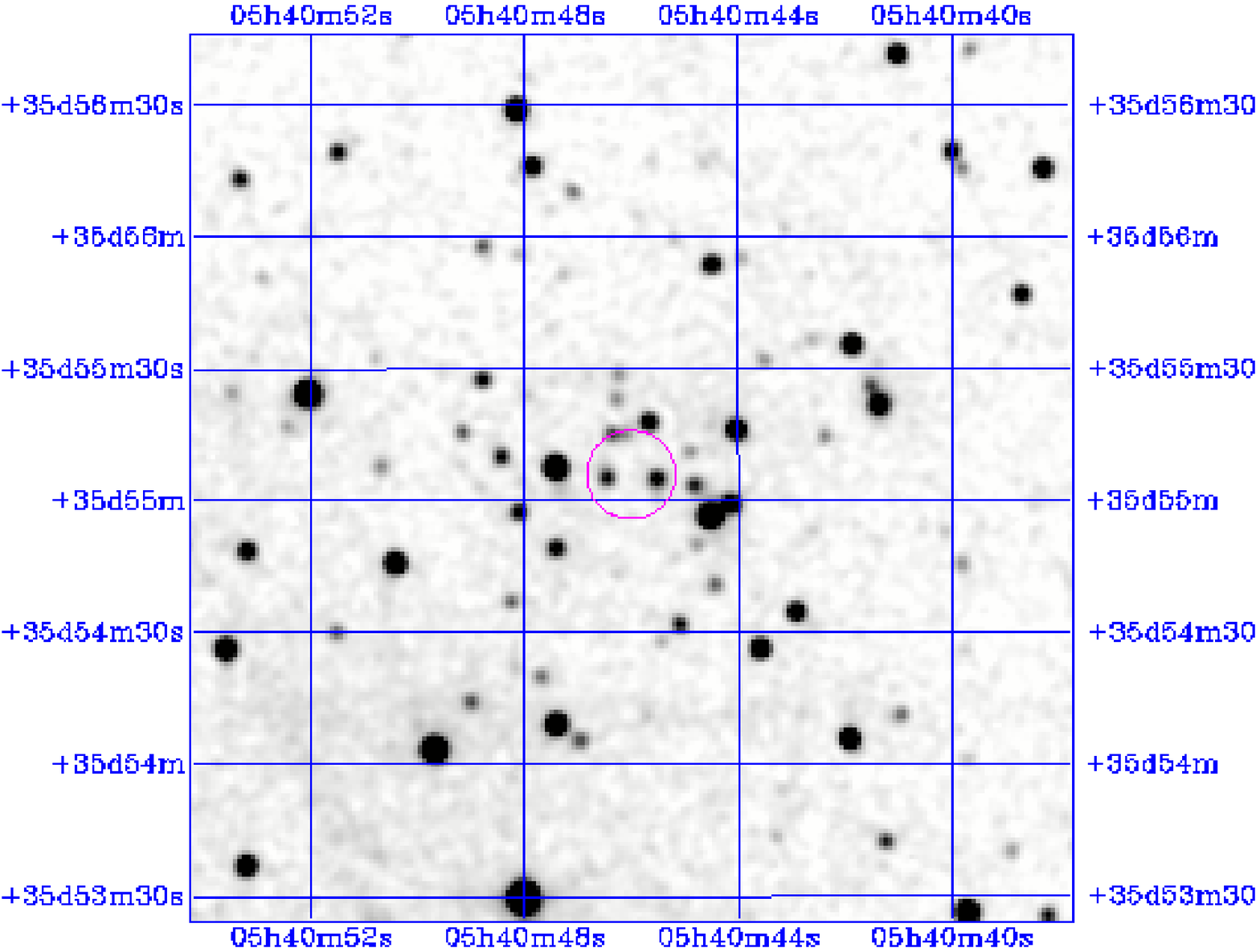}
   \includegraphics[scale=0.32,viewport=0 0 490 460,clip]{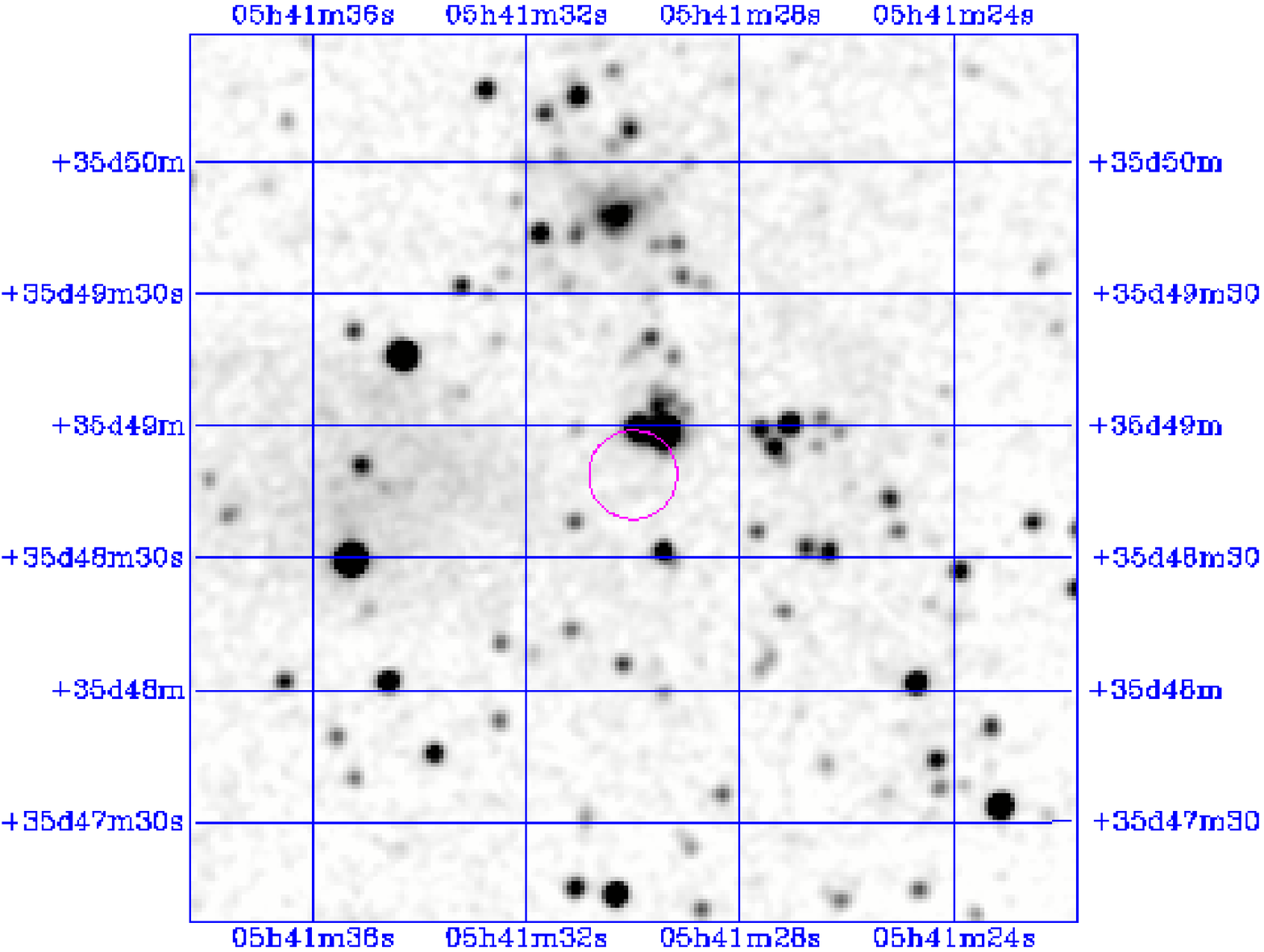}
   \includegraphics[scale=0.32,viewport=0 0 490 460,clip]{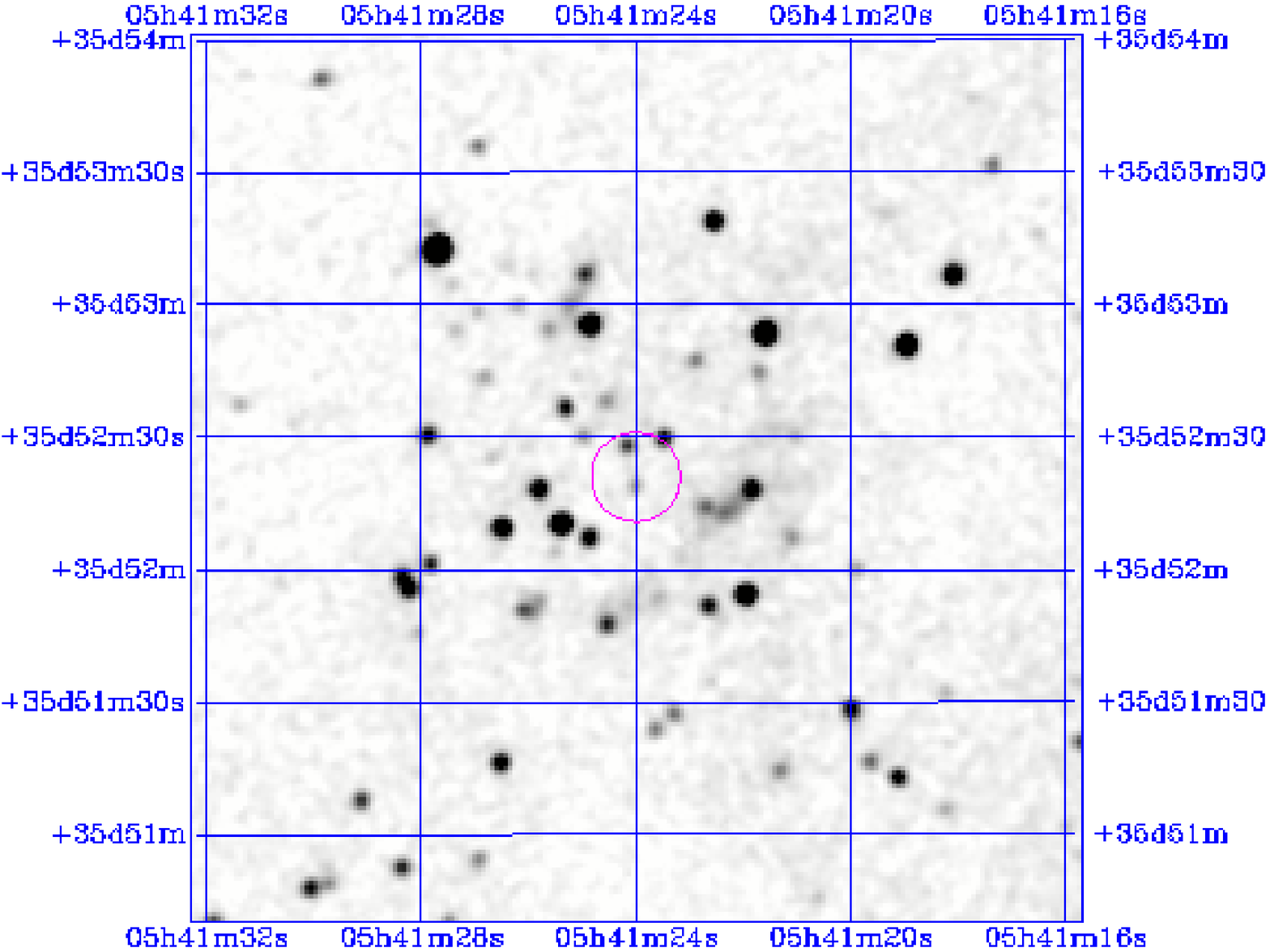}
   \includegraphics[scale=0.32,viewport=0 0 490 460,clip]{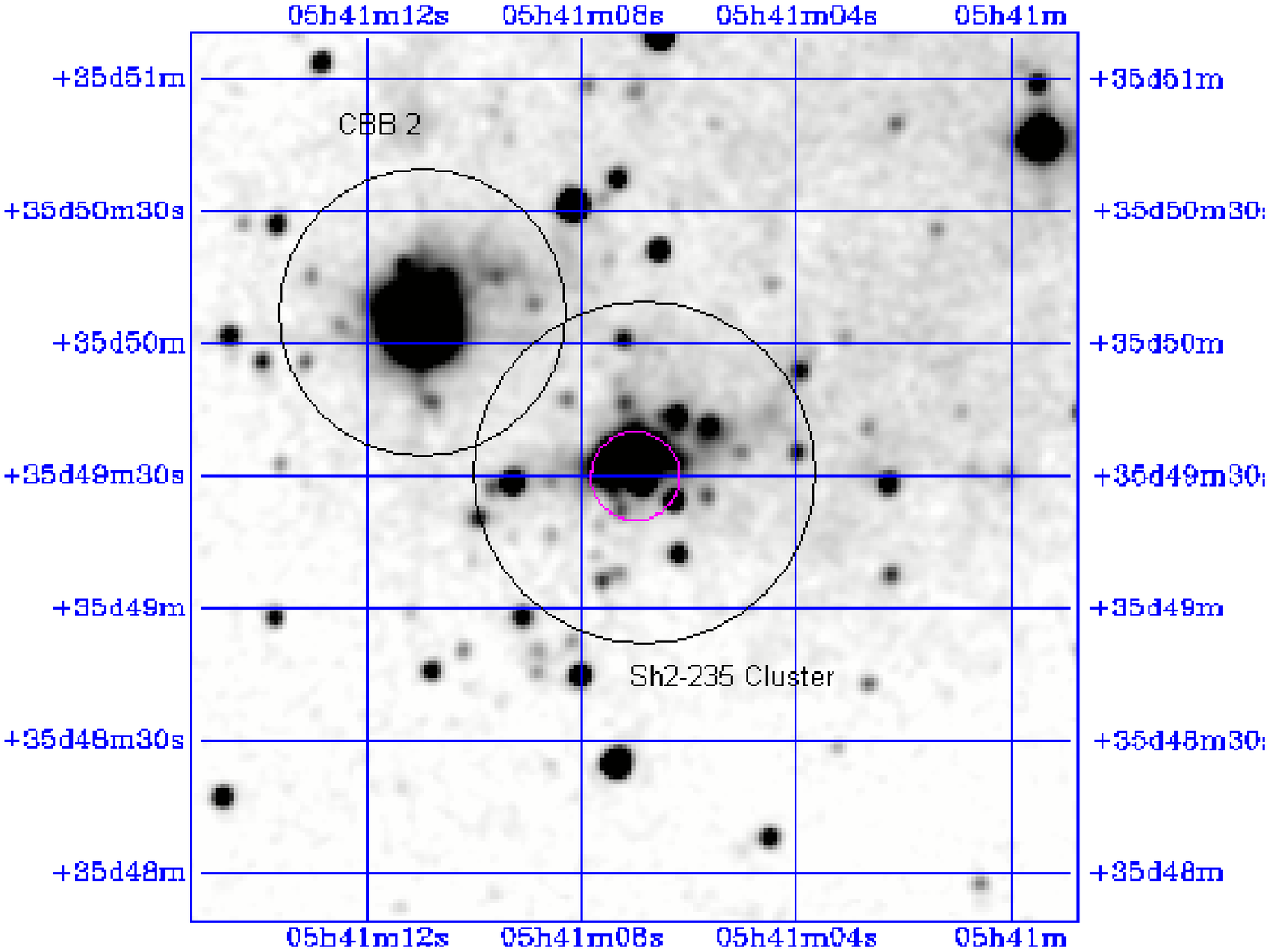}
   \includegraphics[scale=0.32,viewport=0 0 490 460,clip]{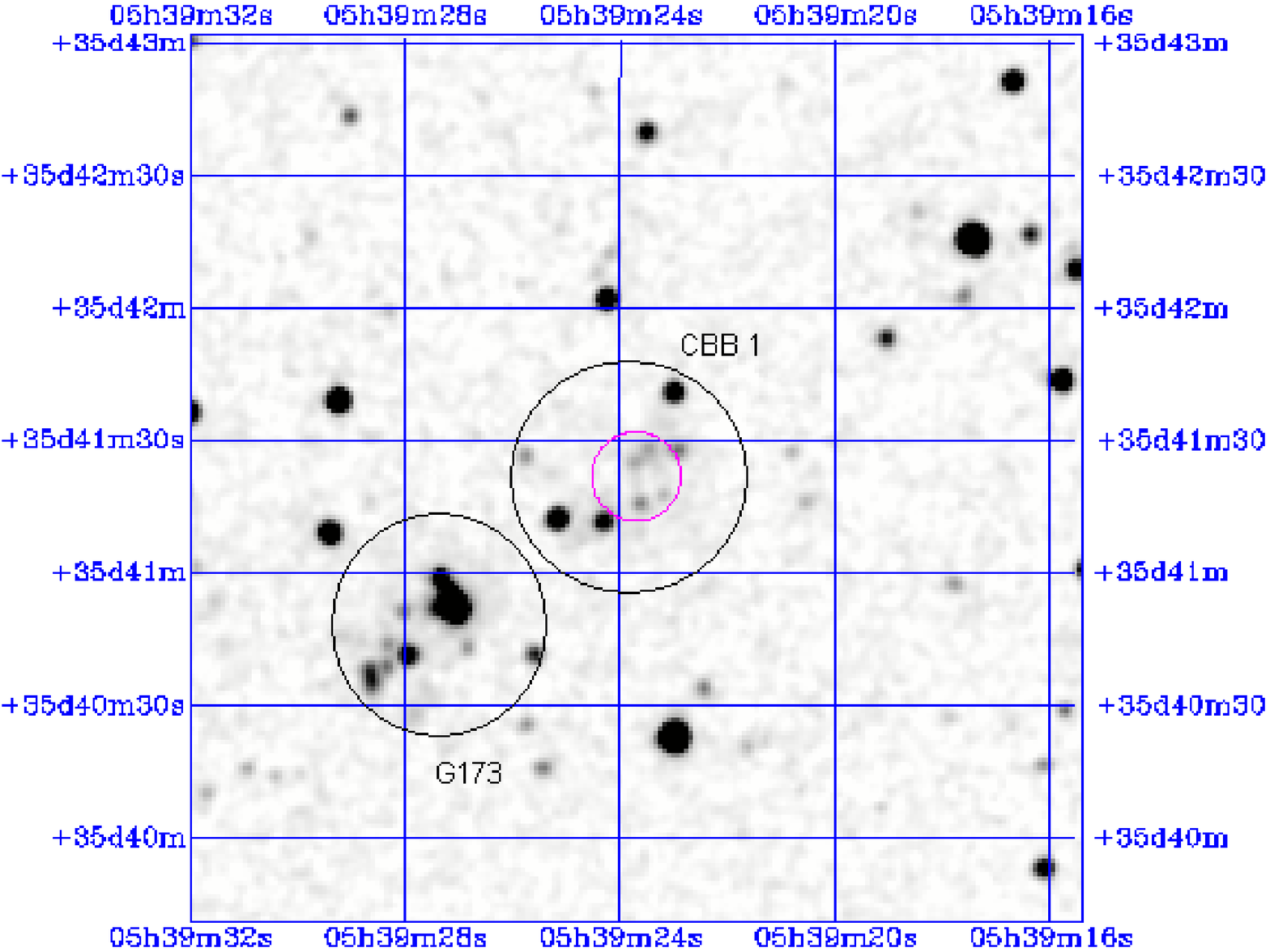}
   \includegraphics[scale=0.32,viewport=0 0 490 460,clip]{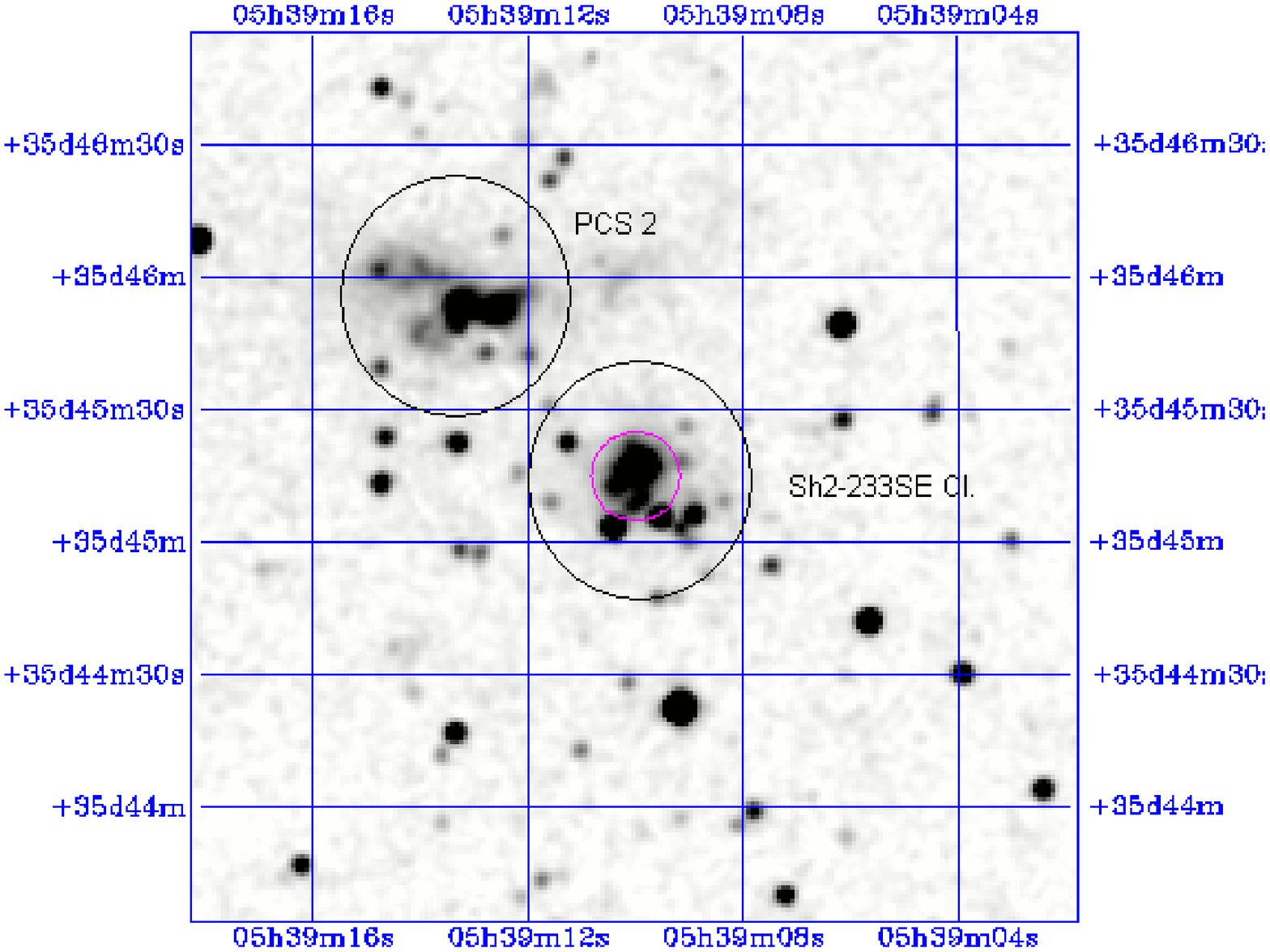}
   \includegraphics[scale=0.32,viewport=0 0 490 460,clip]{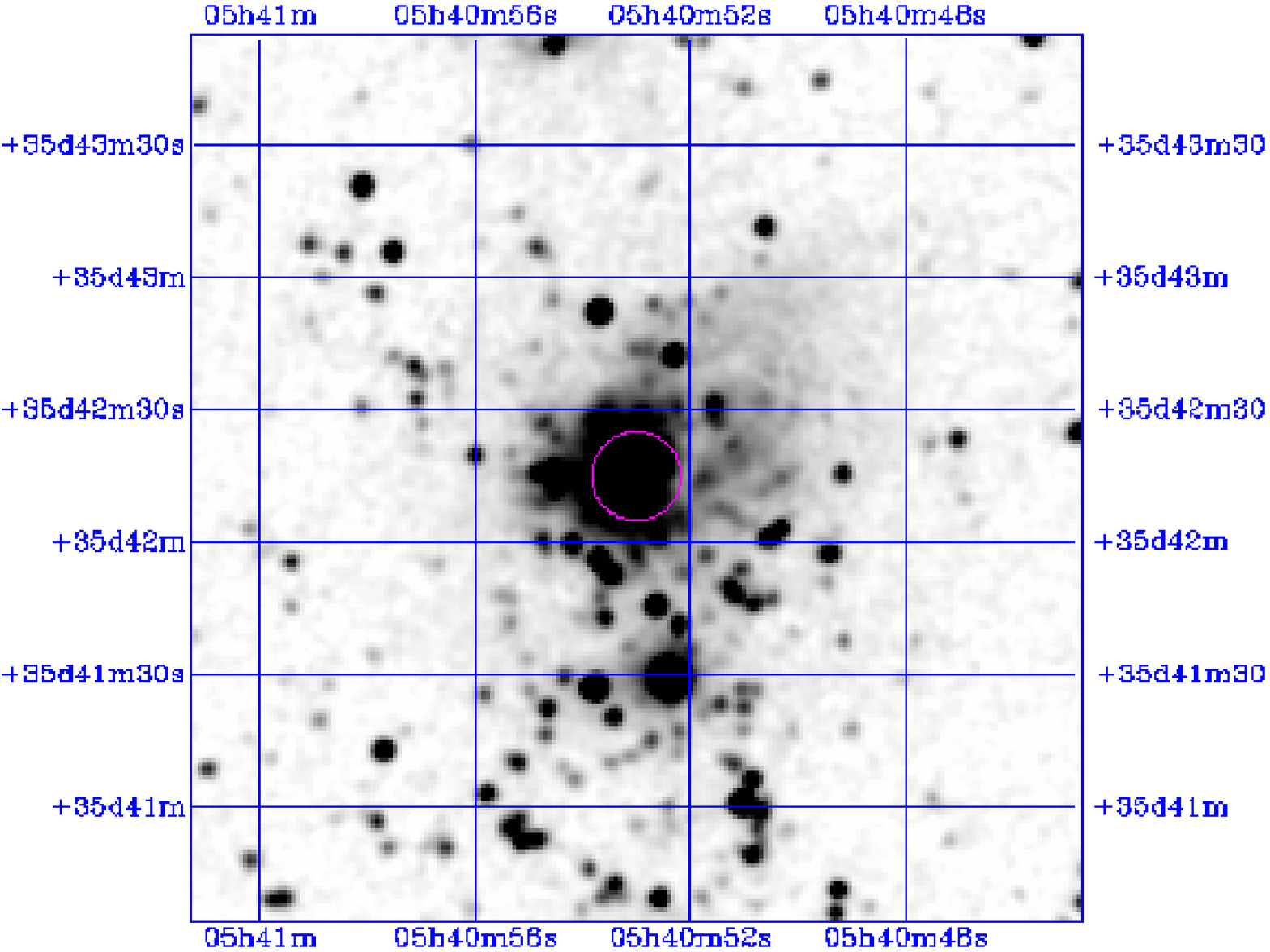}
   \includegraphics[scale=0.32,viewport=0 0 490 460,clip]{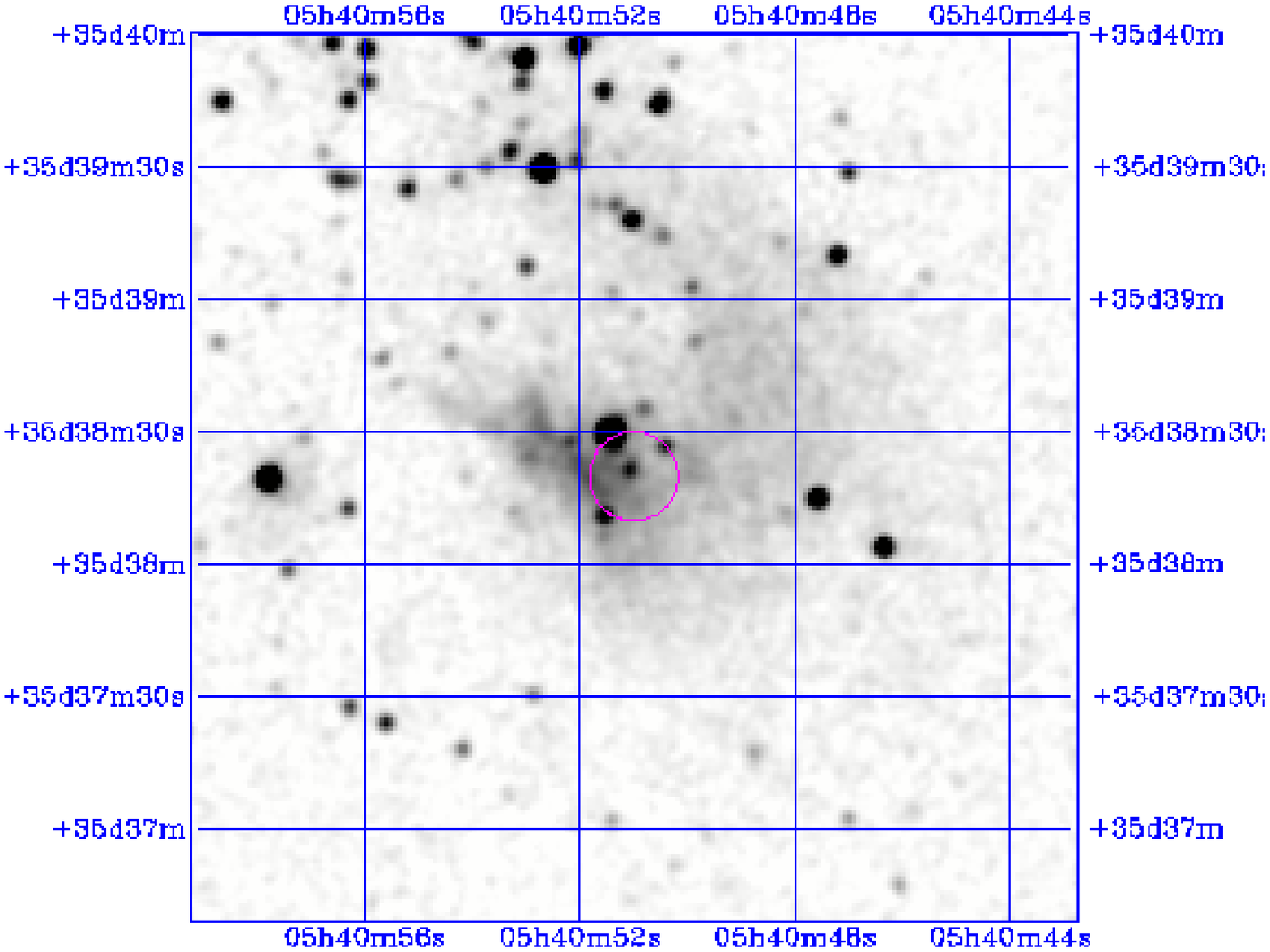}
   \includegraphics[scale=0.32,viewport=0 0 490 460,clip]{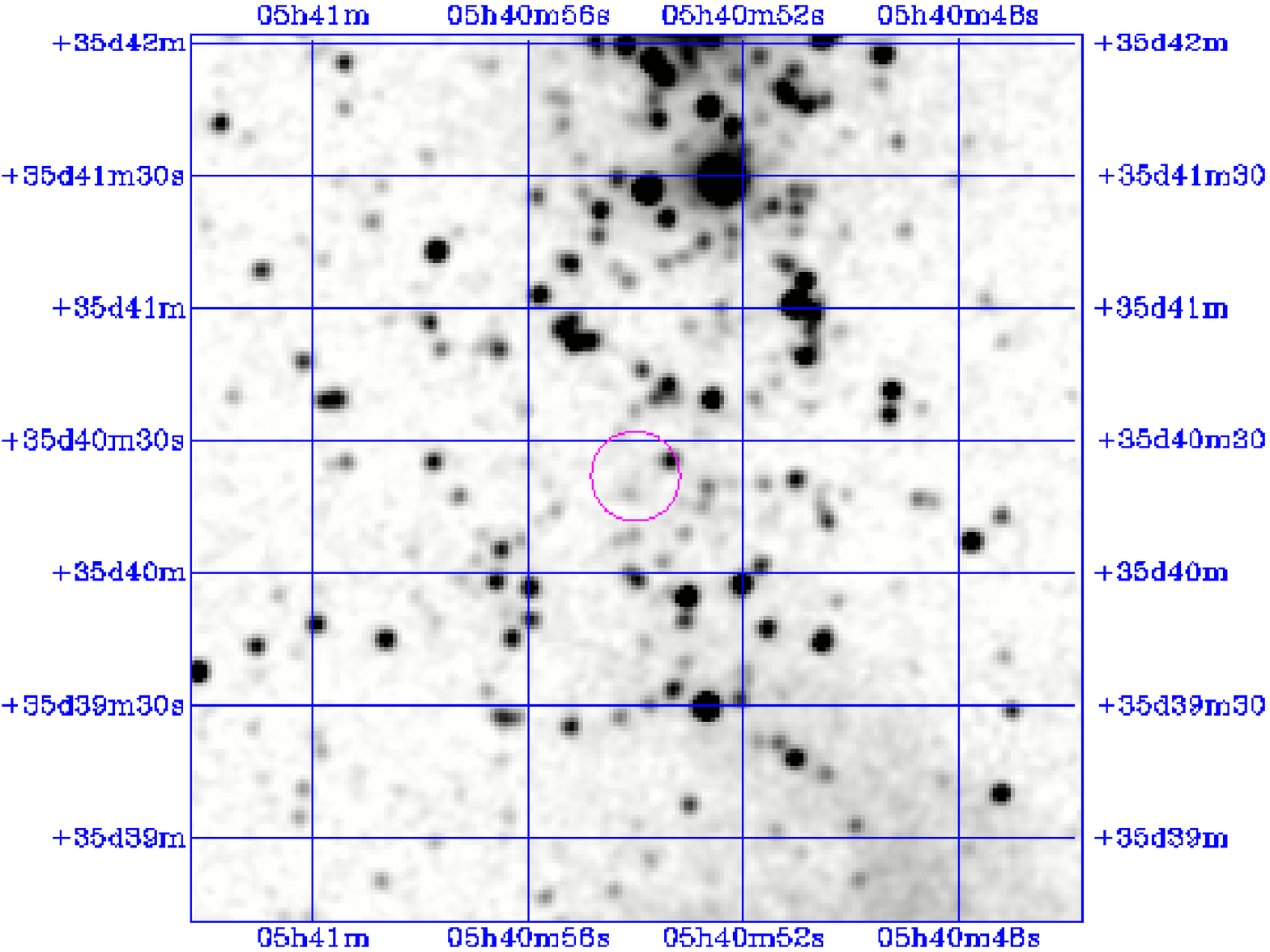}
   \includegraphics[scale=0.32,viewport=0 0 490 460,clip]{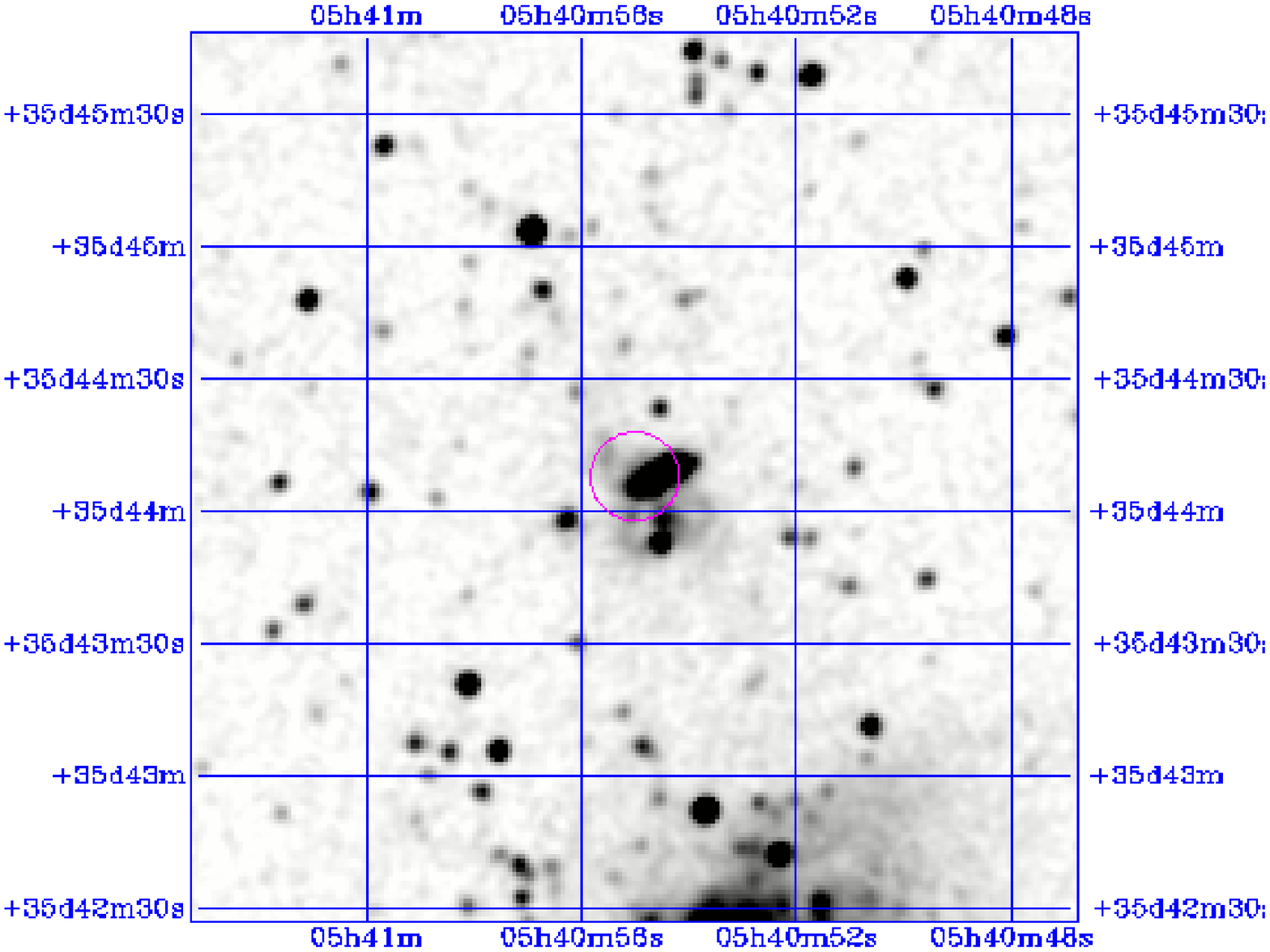}
   \includegraphics[scale=0.32,viewport=0 0 490 460,clip]{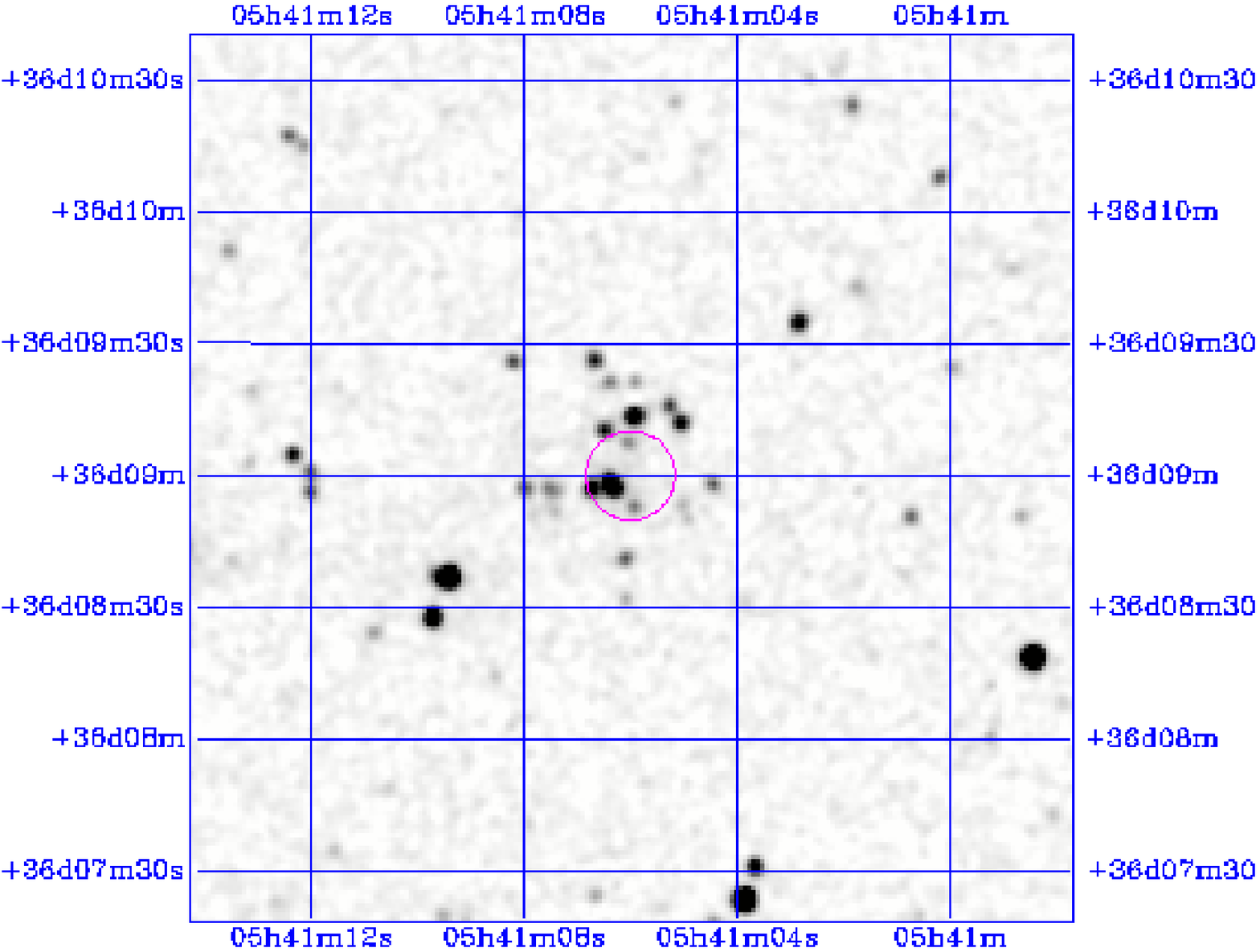}
   \includegraphics[scale=0.32,viewport=0 0 490 460,clip]{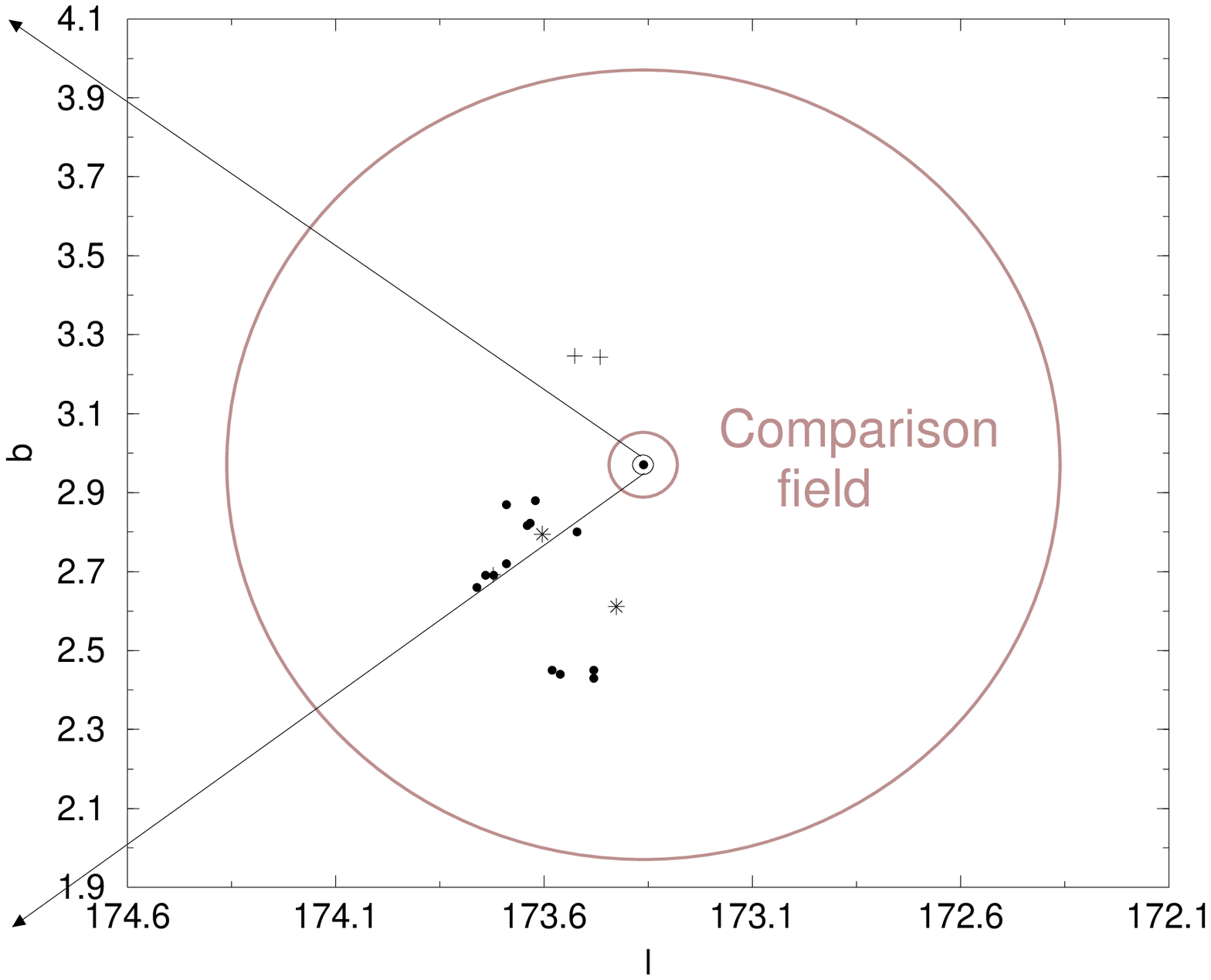}
   \caption[]{First row panels: 2MASS $3'\times3'$ $K_s$ images centred on FSR 784, KKC 11 and Sh2-235E2. Second row: the same for the cluster pairs Sh2-235 Cluster and CBB 2, G173 and CBB 1, PCS 2 and Sh2-233SE Cluster. Third row: Sh2-235B Cluster, BDSB 71, and BDSB 72. Fourth row: same as row three for BDSB 73 and Sh2-232 IR Cluster. The small circle is automatically generated by the 2MASS image tool, surrounding the input coordinates. The large circle in the second row surrounds the main body of each cluster. The last panel in the fourth row shows the large background area (indicated with  outer and inner brown circles) used in the statistical field decontamination of Sh2-232 IR Cluster. Note that it is huge with respect to the cluster area, for statistical purposes. The symbols are the same as in Fig.~\ref{fig:02}.}
   \label{fig:03}
\end{figure*}

\section{Previous studies on the G174+2.5 stellar content}
\label{sec:2}

The H II regions Sh2-235, Sh2-233, Sh2-232, and Sh2-231 have been subject to several studies in different wavelength ranges, but astrophysical parameters for most ECs located in these regions have not been derived so far. The ECs selected for the present analysis are given in Table~\ref{tab2}, where we adopt a chronological criterion for the literature identifications. We show in Fig.~\ref{fig:01} an XDSS $30'\times30'$ R image of the Sh2-235 H II region. In Fig.~\ref{fig:02} we show schematic charts for the nebulosities and clusters, and in Fig.~\ref{fig:03} 2MASS images of the ECs in the $K_S$ band. All objects are quite obscured by interstellar dust and appear to be affected by differential reddening. Being projected close to each other, this sample of ECs may be an example of H II regions with a sequential star formation and the development of a \textit{collect and collapse} scenario. Furthermore,  the detection of far-IR sources, molecular outflows, and $H_{2}O$ masers in previous works indicate on-going star formation. The distance to the Sun estimated for these H II regions is in the range $d_{\odot}=1.0$-$2.3$ kpc \citep{Georgelin75}. Most works use $d_{\odot}=1.8$ kpc for this group of nebulosities. 

Within uncertainties, Sh2-231, Sh2-232, Sh2-233 and Sh2-235 have comparable CO radial velocities \citep{Blitz82}. This indicates
that we are dealing with a large HII-molecular complex with components located essentially at the same distance from the Sun.

\begin{figure}
\resizebox{\hsize}{!}{\includegraphics{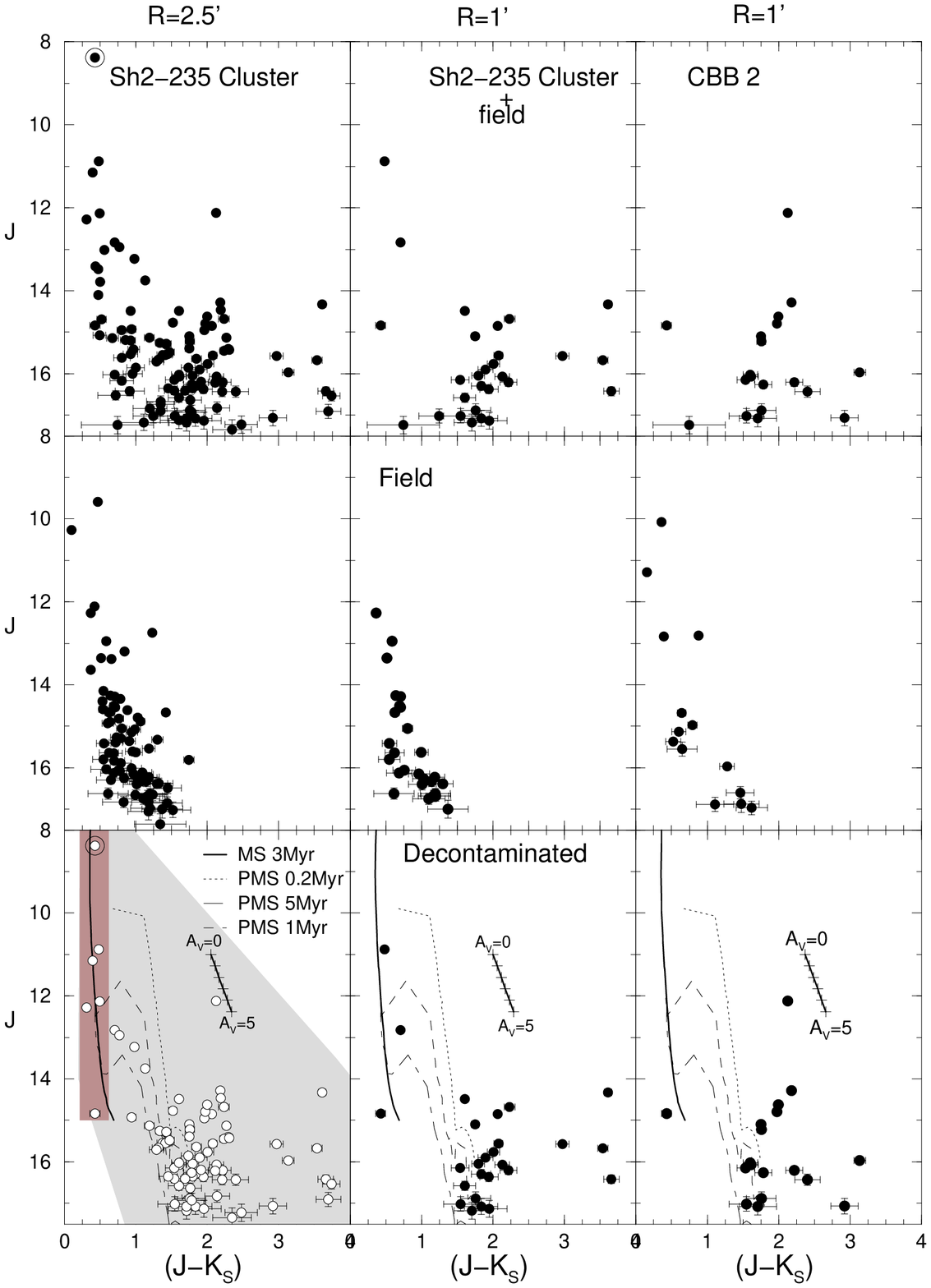}}
\caption[]{2MASS CMDs extracted from the central regions of the cluster pair Sh2-235 Cluster and CBB 2. Top panels: observed CMDs $J\times(J-K_s)$. Middle panels: equal area comparison field. Bottom panels: field star decontaminated CMDs fitted with the Padova isochrone for MS stars and Siess for PMS stars. The colour-magnitude filters used to isolate cluster MS and PMS stars are shown as  shaded regions. We also present the reddening vector for $A_V=0$ to 5.  BD$+35^{\circ}1201$ is shown as a circle around the star in the top and bottom-left panels.}
\label{fig:04}
\end{figure}

\begin{figure}
\resizebox{\hsize}{!}{\includegraphics{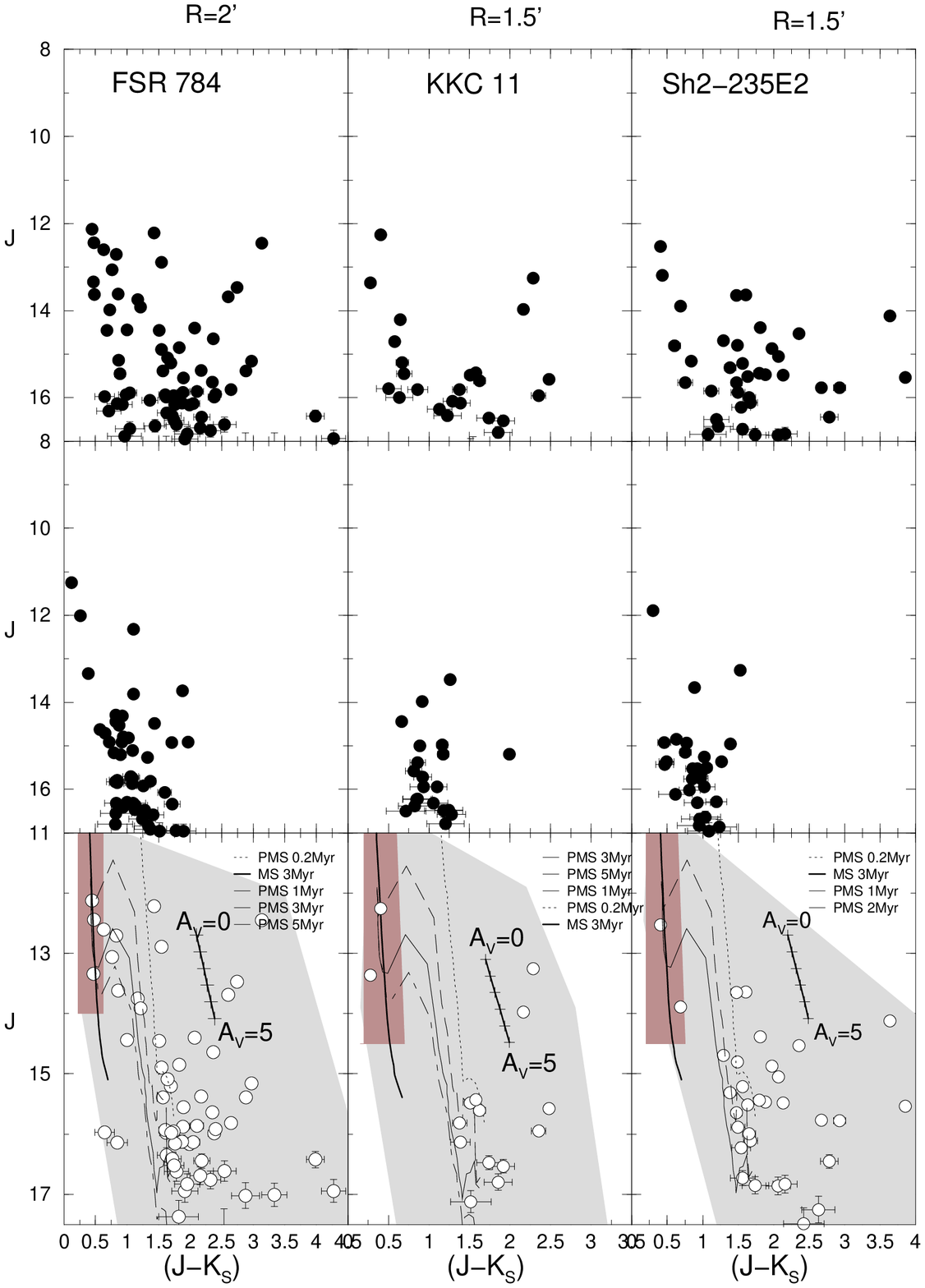}}
\caption[]{2MASS CMDs extracted from the central regions of FSR 784, KKC 11 and Sh2-235 E2, respectively. Top panels: observed CMDs $J\times(J-K_s)$. Middle panels: equal area comparison field. Bottom panels: field star decontaminated CMDs fitted with the Padova isochrone for MS stars and \citet{Siess00} for PMS stars. The colour-magnitude filters used to isolate cluster MS and PMS stars are shown as  shaded regions. We also present the reddening vector for $A_V=0$ to 5.}
\label{fig:05}
\end{figure}

\subsection{Sh2-235 and surroundings}

Sh2-235 is the most prominent H II region in this group. It is a diffuse optical H II region excited  by a star of spectral type O9.5 V (BD$+35^{\circ}1201$). 
\citet{Allen05} identify two clusters associated with the Sh2-235 H II region, Sh2-235 Cluster and  KKC 11. \citet{Kumar06} add to these objects the cluster Sh2-235 East2. \citet{Kirsanova08} conclude that these objects are still embedded in dense clumps of the parental molecular cloud $G174+2.5$. They also argue that the Sh2-235 Cluster and Sh2-235 E2 probably started the primordial gas expulsion, but KKC 11 is less evolved.

\begin{figure}
\resizebox{\hsize}{!}{\includegraphics{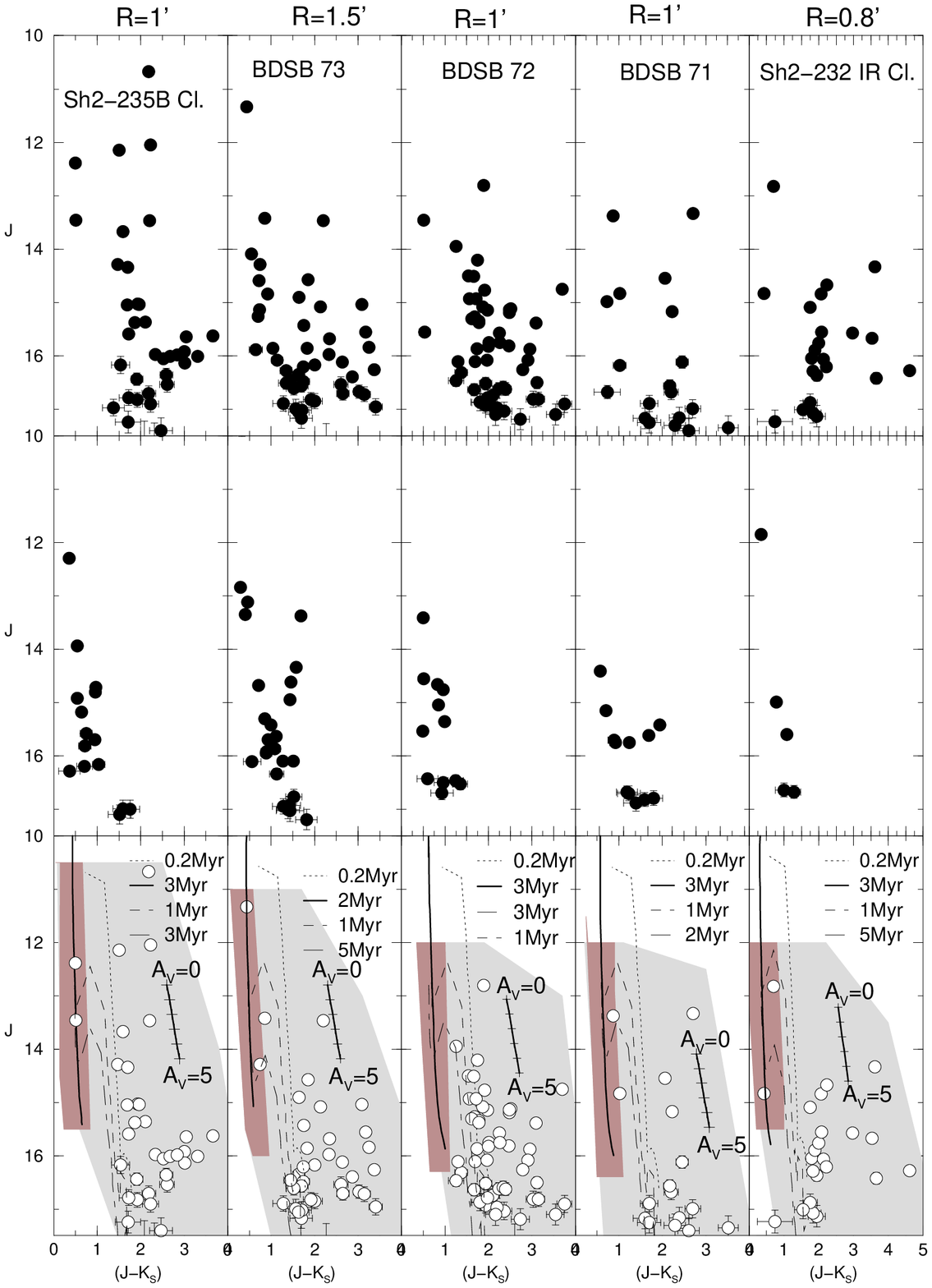}}
\caption[]{Same as Fig.~\ref{fig:05} for Sh2-235 B, BDSB 73, BDSB 72, BDSB 71 and Sh2-232 IR Cluster, respectively.}
\label{fig:06}
\end{figure}

\begin{figure}
\resizebox{\hsize}{!}{\includegraphics{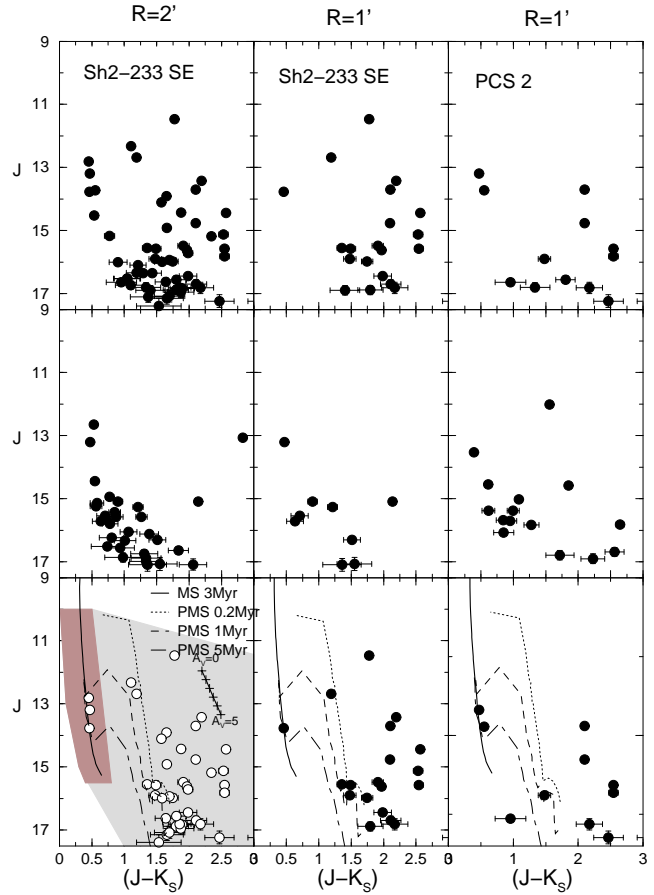}}
\caption[]{Same as Fig.~\ref{fig:04} for the cluster pair PCS 2 and Sh2-233SE.}
\label{fig:07}
\end{figure}

South-west of the Sh2-235 are four small nebulae named GGD 6 (also known as RNO 52), Sh2-235 A (GM1-G6), Sh2-235 B (BFS 46 or GM1-G5) and Sh2-235 C (GGD 5 or BFS 47).
Sh2-235 A and Sh2-235 B are located about $10\arcmin$ (5.6 pc) south of Sh2-235, and separated by $\approx40\arcsec$ (0.35 pc) \citep{Blitz82}. Sh2-235 A is a compact H II region of $\approx20\arcsec$ (0.17 pc) in diameter \citep{Felli97}. For \citet{Boley09} the exciting star of Sh2-235 B is an early-type Herbig Be star of spectral type B1V.
\citet{Hodapp94} found a cluster in Sh2-235 B, which was confirmed by \citet{Wang97}, \citet{Allen05} and \citet{Kirsanova08}. 
Sh2-235 C is a small H II region located about $3.5\arcmin$ south of Sh2-235 B \citep{Felli04}.
It coincides with an optical nebula and is excited by a B0.5 star, presenting a Herbig-Haro object and a partial shell morphology \citep{Felli97}. \citet{Bica03} identify other three ECs in these nebulosities, BDSB 71, 72 and 73. 

Based on the spatial distribution of clusters and kinematics of molecular gas \citet{Kirsanova08} point out, that the expansion of Sh2-235 would be responsible for cluster formation in that area. They suggest that sequential star formation is triggered by a combination between compression of pre-existing dense clumps by the shock wave, and the  \textit{collect and collapse} scenario. However, the clusters in Sh2-235 A, B and C appear to be embedded in primordial gas, and star formation cannot be triggered by the expansion of the  Sh2-235 ionisation front \citep[][]{Lafon83}. On the other hand, \citet{Tokunaga79} suggest that the linear disposition of these star forming regions may be the result of \textit{collect and collapse} scenario. 

\subsection{Sh2-231, Sh2-232 and Sh2-233}

Sh2-232 is an extended H II region with $\approx40\arcmin$ diameter (Fig.~\ref{fig:02}), excited by B stars. 
\citet{Hodapp94} observed two nebulae in Sh2-233 and noticed a young infrared EC, as well as a probable Herbig-Haro object.
\citet{Porras00} identified two clusters in this region, PCS 2, located around the IRAS $05358+3543$ source and Sh2-233 SE at a separation of $1\arcmin$ (0.5 pc) from each other for an adopted distance to the Sun of 1.8 kpc. However, \citet{Chan95} estimated a distance of  $2.3\pm0.7$ kpc, based on spectral types of stars in the H II region. \citet{Porras00} detected 92 stars in JHK, but the those associated to PCS 2 is 20 and to Sh2-233 SE is 15 stars. The average extinction was $A_V=8.44\pm4.77$ for Sh2-233 SE and $A_V=15.06\pm3.48$ for PCS 2. The estimated ages were 6 Myr for field stars, 3 Myr for Sh2-233 SE and less than 2 Myr for PCS 2.  \citet{Mao04}, based on cluster members \citep{Porras00}, obtained a stellar mass of $10.7\,M_\odot$ for  PCS 2 with a star formation efficiency (SFE) of $\sim4\%$, and for Sh2-233 SE a mass of $38.4\,M_\odot$ with a SFE of $\sim47\%$. They suggest that the SFE of PCS 2 was underestimated because of the upper mass completeness limit of $1\,M_\odot$ estimated for PCS 2, and point out that the SFE increases with time while the cluster is embedded in a molecular cloud. The large SFE difference of the two clusters indicates that Sh2-233 SE is more evolved than PCS 2. \citet{Snell90} identified a CO outflow with two lobes centred on IRAS $05358+3543$. \citet{Jiang01} named Sh2-233SE and PCS 2 as S233 A and S233 B, respectively. They argue that Sh2-233SE is less embedded than PCS 2. Recently, \citet{Yan10} derived ages of $0.3$, $0.5$, and $1.5$ Myr and masses of $45$, $30$, and $107\,M_\odot$ for Sh2-233 SE Cluster, PCS2 and field stars, respectively. 

The nebula G$173.58+2.45$ shows multiple outflow sources, typical of regions presenting young stellar objects (YSOs) \citep{Wouterloot89, Shepherd96, Shepherd02, Varricatt05}. The  most extended jet appears to be associated with a binary system near the centre of the G173.58+2.45 Cluster (hereafter G173). We discovered an additional EC, CBB 1, at an angular distance of $1\arcmin$ from G173.

\section{2MASS photometry}
\label{sec:3}

2MASS\footnote{The Two Micron All Sky Survey, available at \textit{www..ipac.caltech.edu/2mass/releases/allsky/}} photometry \citep{Skrutskie06} in the $J$, $H$ and $K_{s}$ bands was extracted in concentric regions centred on the coordinates of the ECs (Table~\ref{tab4}) using VizieR\footnote{http://vizier.u-strasbg.fr/viz-bin/VizieR?-source=II/246.}. Large extraction areas are essential to build RDPs (Sect.~\ref{sec:4}) with a high resulting contrast relative to the background, and for a consistent field star decontamination (Sect.~\ref{sec:3.1}). 

\subsection{Field-star decontamination}
\label{sec:3.1}

\begin{figure}
\resizebox{\hsize}{!}{\includegraphics{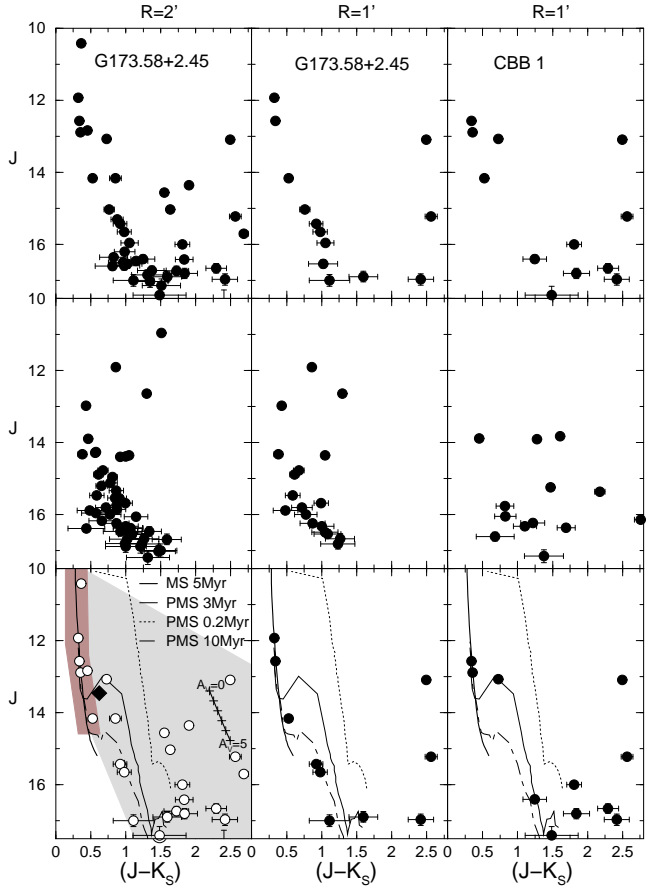}}
\caption[]{Same as Fig.~\ref{fig:04} for the cluster pair G173 and CBB 1. The diamond is the IRAS 05361+3539 source.}
\label{fig:08}
\end{figure}

Field stars are conspicuous in the observed CMDs of the present ECs (Figs.~\ref{fig:04} to \ref{fig:08}). To uncover the intrinsic CMD morphology from the field stars, we apply a field-star decontamination procedure. The algorithm deals statistically the relative number-densities of probable cluster and field stars in cubic CMD cells that have axes along the $J$, $(J-H)$ and $(J-K_{s})$ axes. These are the colours that provide the maximum discrimination among CMD sequences for star clusters of different ages \citep[e.g.][]{Bonatto04}. 

The algorithm (i) divides the range of magnitude and colours of a given CMD into a 3D grid, (ii) computes the expected number-density of field stars in each cell based on the number of comparison field stars (within  $1\sigma$ Poisson fluctuation) with magnitude and colours compatible with those of the cell, and (iii) subtracts from each cell a number of stars that corresponds to the number-density of field stars measured within the same cell in the comparison field. Consequently, this method is sensitive to local variations in field star contamination in magnitude and colours. Cell dimensions are $\Delta{J}=1.0$, and  $\Delta(J-H)={\Delta(J-K_{s})}=0.2$, which provide sufficient star-count statistics in individual cells and preserve the morphology of the CMD evolutionary sequences. The dimensions of the colour/magnitude cells can be subsequently changed so that the total number of stars subtracted in the whole cluster area matches the expected one ($1\sigma$ Poisson fluctuation). We gave a brief description of the field-star decontamination procedure. For details see \citet{Bonatto07a} and \citet{Bica08}.

\begin{table*}
{\footnotesize
\begin{center}
\caption{Derived fundamental parameters for the embedded clusters.}
\renewcommand{\tabcolsep}{4.4mm}
\renewcommand{\arraystretch}{1.3}
\begin{tabular}{lrrrrrrrrr}
\hline
\hline
Cluster&$\alpha(2000)$&$\delta(2000)$&$\ell$&$b$&$A_V$&Age&$d_{\odot}$&$R_{GC}$\\
&(h\,m\,s)&$(^{\circ}\,^{\prime}\,^{\prime\prime})$&$(^{\circ})$&$(^{\circ})$&(mag)&Myr&(kpc)&(kpc)\\
($1$)&($2$)&($3$)&($4$)&($5$)&($6$)&($7$)&($8$)&($9$)\\
\hline
KKC\,11 &5:41:30&35:48:49&$173.685$&$2.865$&$5.5\pm1.0$&$3\pm2$&$2.2\pm0.5$&$9.4\pm0.2$\\
FSR\,784 &5:40:46&35:55:06&$173.517$&$2.794$&$4.0\pm1.0$&$3\pm2$&$2.4\pm0.5$&$9.6\pm0.2$\\
Sh2-235\,E2 &5:41:24&35:52:21&$173.624$&$2.878$&$4.0\pm1.0$&$3\pm2$&$2.1\pm0.5$&$9.3\pm0.2$\\
Sh2-235\,Cl. &5:41:07&35:49:30&$173.634$&$2.805$&$3.8\pm1.0$&$5\pm2$&$2.0\pm0.6$&$9.2\pm0.2$\\
BDSB\,73 &5:40:55&35:44:08&$173.688$&$2.723$&$3.8\pm1.0$&$3\pm2$&$2.1\pm0.5$&$9.3\pm0.5$\\
Sh2-235B Cl.&5:40:53&35:42:15&$173.712$&$2.701$&$3.8\pm1.0$&$3\pm2$&$1.9\pm0.5$&$9.1\pm0.5$\\
BDSB\,72 &5:40:54&35:40:22&$173.740$&$2.687$&$3.8\pm1.0$&$3\pm2$&$2.1\pm0.5$&$9.3\pm0.5$\\
BDSB\,71 &5:40:51&35:38:20&$173.763$&$2.660$&$3.8\pm1.0$&$3\pm2$&$2.0\pm0.4$&$9.2\pm0.4$\\
Sh2-232\,IR &5:41:06&36:09:00&$173.356$&$2.973$&$5.0\pm1.2$&$3\pm2$&$1.9\pm0.5$&$9.1\pm0.5$\\
PCS\,2 &5:39:13&35:45:53&$173.481$&$2.446$&$3.5\pm1.0$&$3\pm2$&$2.2\pm0.5$&$9.4\pm0.5$\\
Sh2-233\,SE &5:39:10&35:45:15&$173.484$&$2.432$&$3.5\pm0.8$&$3\pm2$&$2.2\pm0.5$&$9.4\pm0.5$\\
$G173$\,Cl. &5:39:28&35:40:43&$173.581$&$2.443$&$3.5\pm0.9$&$5\pm3$&$2.5\pm0.2$&$9.8\pm0.2$\\
CBB\,1 &5:39:23&35:41:22&$173.563$&$2.434$&$3.5\pm0.9$&$5\pm3$&$2.5\pm0.2$&$9.8\pm0.2$\\
CBB\,2 &5:41:11&35:50:10&$173.632$&$2.822$&$3.8\pm1.0$&$3\pm2$&$2.0\pm0.6$&$9.2\pm0.2$\\
\hline
\end{tabular}
\begin{list}{Table Notes.}
\item Cols. 2 to 5: Optimised central coordinates; Col. 6: reddening in the cluster's central region. Col. 7: age, from 2MASS photometry. Col. 8: distance to the Sun. Col. 9: $R_{GC}$ calculated with $R_{\odot}=7.2$ kpc \citep{Bica06} as the distance of the Sun to the Galactic centre. The ECs CBB 1 and CBB2 were discovered in the present work.
\end{list}
\label{tab4}
\end{center}
}
\end{table*}

\begin{figure}
\resizebox{\hsize}{!}{\includegraphics{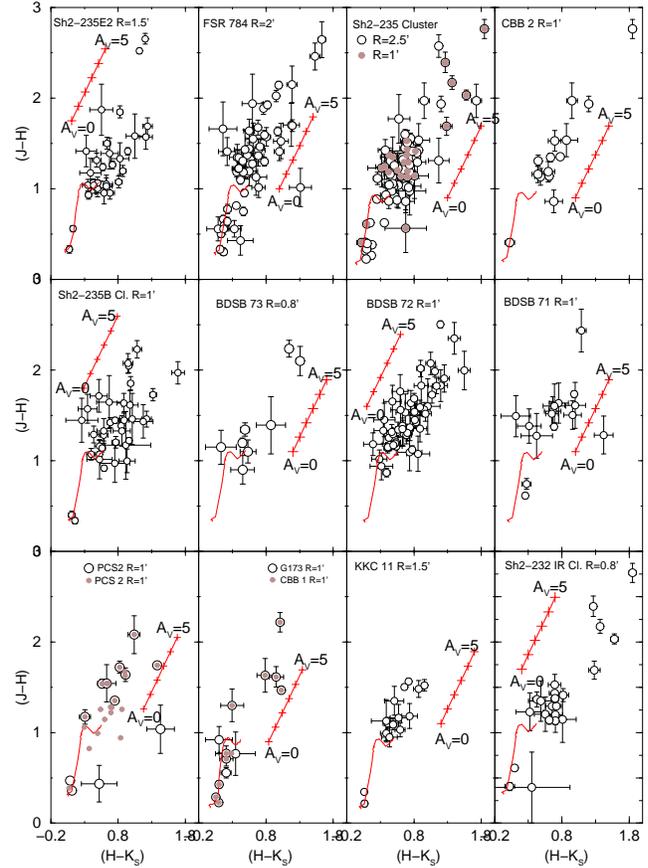}}
\caption[]{Colour-colour diagrams for the decontaminated photometry. \citet{Siess00} isochrones and reddening vectors are compared with the PMS distribution.}
\label{fig:09}
\end{figure}

\subsection{Fundamental parameters}
\label{sec:3.2}

Usually, decontaminated CMDs of ECs are characterised by a poorly-populated MS and a significant number of PMS stars. PMS stars are usually located in the same CMD region occupied by faint red field stars and consequently one is not able to distinguish them, considering only CMD properties. In this context, the field-star decontamination procedure is essential for the identification of cluster evolution sequences. We adopt isochrones from the Padova group with solar-metallicity \citep{Girardi02} computed with the 2MASS $J$, $H$ and $K_{s}$ filters, and the PMS tracks of \citet{Siess00} fitted in $J\times(J-H)$ and $J\times(J-K_s)$ decontaminated CMDs (Figs. ~\ref{fig:04} to \ref{fig:08}) to derive the ECs' fundamental parameters.
Because of the poorly-populated MSs, the 2MASS photometric uncertainties for lower sequences, the large population of PMS stars and differential reddening, we decided for the direct comparison of isochrones with the decontaminated CMD morphology. We carry out eye fits, taking the combined MS and PMS stellar distributions as constraint, allowing for differential reddening and photometric uncertainties. The cluster sample presents a significant fraction of stars redder than the youngest PMS isochrone. Probably, the differential reddening contributed to this $(J-K_s)$ excess towards red colours, but part of it can be intrinsic and related to their evolutionary stage. The youngest stars are the reddest and with evolution their colours become bluer and move towards the MS \citep{Lada92b}. Most of these redder stars have infrared excesses owing to disks.
The isochrone fit gives the observed distance modulus $(m-M)_{J}$ and reddening $E(J-H)$, which converts to $E(B-V)$ and $A_{V}$ with the relations $A_{J}/{A_{V}}=0.276$, $A_{H}/{A_{V}}=0.176$, $A_{K_{s}}/{A_{V}}=0.118$, $A_{J}=2.76\times{E(J-H)}$  and $E(J-H)=0.33\times{E(B-V)}$ \citep{Dutra02}. We estimate $A_{V}$ assuming a constant total-to-selective absorption ratio $R_{V}=3.1$ and adopt the Sun's distance to the Galactic centre $R_{\odot}=7.2\,kpc$ \citep{Bica06}.
The resulting $A_{V}$, age, $d_{\odot}$ and $R_{GC}$ values are given in cols. 4 to 7 of Table ~\ref{tab4}, respectively.
We also present in the field subtracted CMDs the reddening vector for $A_V=0$ to 5.

In Fig.~\ref{fig:04} we show the CMDs of the cluster pair Sh2-235 ($R=2.5'$ and $R=1'$) and CBB 2 ($R=1'$). The aim of this procedure is to isolate the population of each EC and discard the possibility that we may be working with a single object.
In the top-panel we present the $J\times(J-K_S)$ CMDs extracted for Sh2-235 cluster and CBB 2. In the middle panels we show the background field corresponding to a ring with the same area as the central region. In the bottom panels we give the field star decontaminated CMDs. For the ensemble we get $A_V=3.77\pm1.0$ mag, and a distance to the Sun $d_{\odot}=2.0\pm0.1$ kpc, which agrees with previous values \citep{Georgelin73}. 

Fig.~\ref{fig:05} presents the $J\times(J-K_S)$ CMDs for FSR 784, KKC 11 and Sh2-235E2. These objects are very young cluster characterised by the presence of a poorly-populated MS and a larger number of PMS that appears to be very reddened. These features are absent in the comparison field. The gap between the MS and PMS is characteristic of young clusters. The combination of MS and PMS shows that these objects are ECs.

We show in Fig. ~\ref{fig:06} the CMDs of the ECs Sh2-235 B, BDSB 73, BDSB 72, BDSB 71 and Sh2-232 IR Cluster. These objects present a poorly-populated MS and a rather populous PMS. 

Figs.~\ref{fig:07} and ~\ref{fig:08} show the CMDs of cluster pairs. Fig.~\ref{fig:07} shows the CMDs of the pair PCS 2 ($R=1'$) and Sh2-233 SE Cluster ($R=2'$ and $R=1'$). Fig.~\ref{fig:08} presents the CMDs of the pair CBB 1 ($R=1'$) and G173.58 Cluster ($R=2'$ and $R=1'$). Both objects present CMDs characteristic of very young clusters. The pair separations are small and have a number of stars in common. We do not discard the possibility that these two objects will merge. The $d_{\odot}=2.2\pm0.5$ kpc for Sh2-233 SE agrees with $d_{\odot}=2.3\pm0.7$ estimated by \citet{Chan95}. The diamond in the CMD of the G173 is the IRAS 05361+3539 source (Sect. \ref{sec:2}), which appears to be a PMS star approaching the MS.

We estimate $A_{V}$ for BDSB 71 and Sh2-233SE Cluster assuming a constant total-to-selective absorption ratio $R_{V}=5.0$, which is typical of dense clouds \citep{Cardelli89}. BDSB 71 presents $A_{V}=6.0\pm1.6$ mag for a distance from the Sun of $d_{\odot}=1.3\pm0.7$ kpc. For Sh2-233SE Cluster the values are $A_{V}=6.4\pm1.6$ mag and $d_{\odot}=1.6\pm0.5$ kpc.

\begin{figure*}
\begin{minipage}[b]{0.50\linewidth}
\includegraphics[width=\textwidth]{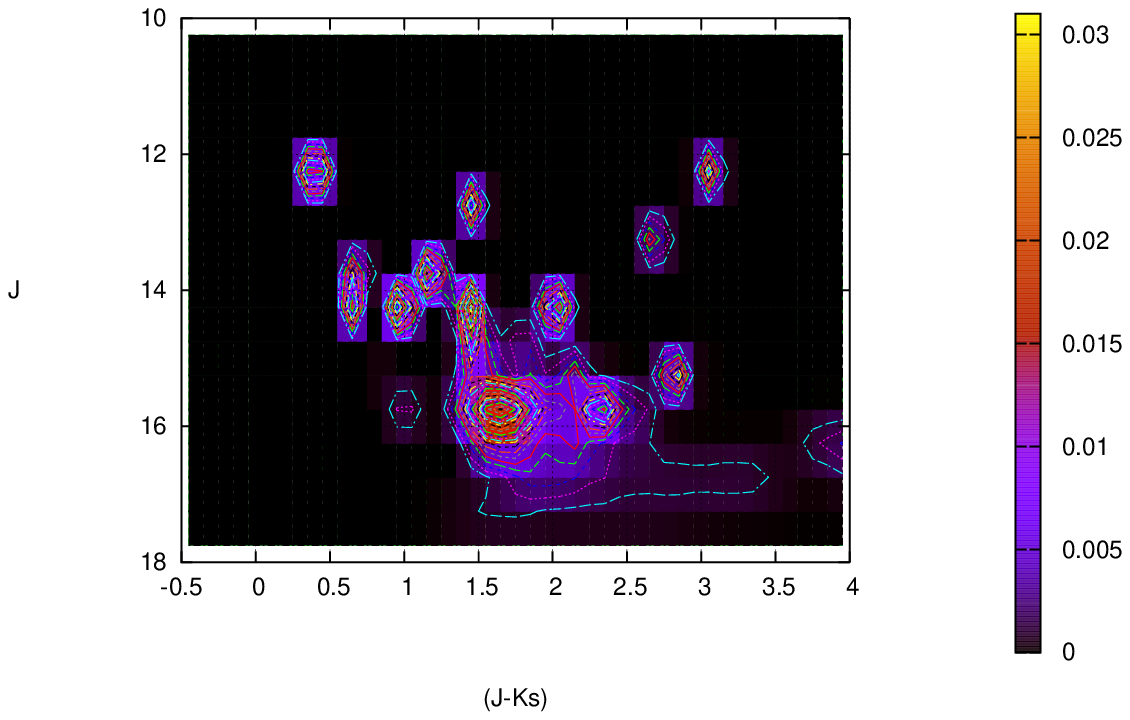}
\end{minipage}\hfill
\begin{minipage}[b]{0.50\linewidth}
\includegraphics[width=\textwidth]{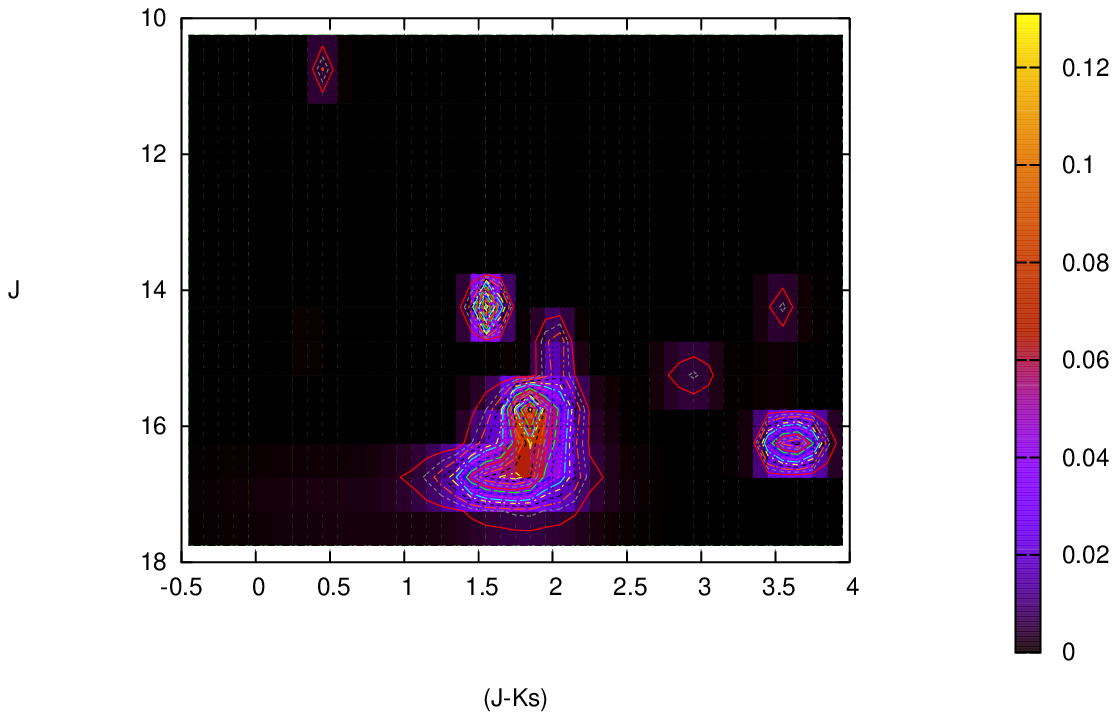}
\end{minipage}\hfill
\begin{minipage}[b]{0.50\linewidth}
\includegraphics[width=\textwidth]{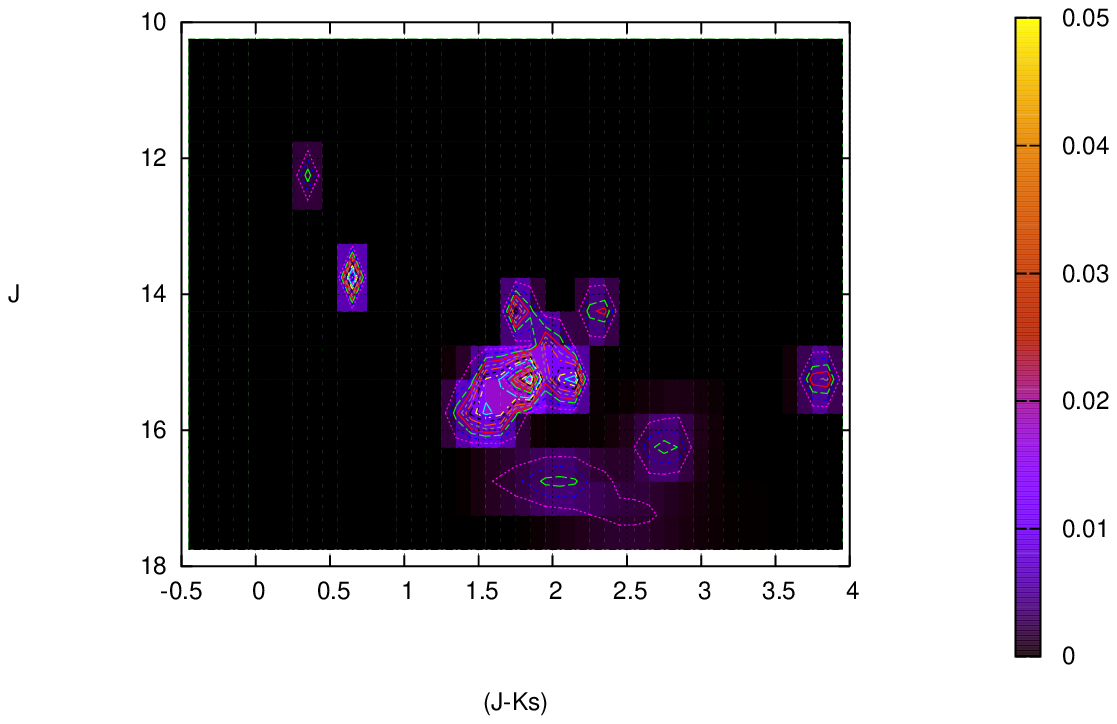}
\end{minipage}\hfill
\begin{minipage}[b]{0.50\linewidth}
\includegraphics[width=\textwidth]{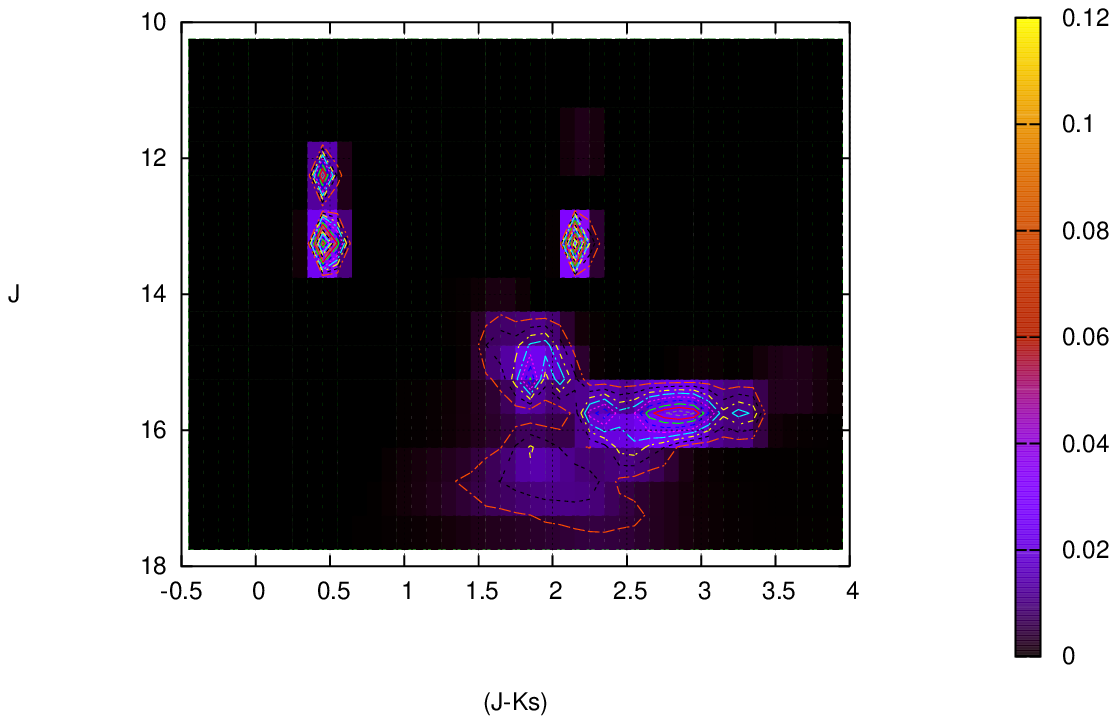}
\end{minipage}\hfill
\caption[]{Number-density (stars $arcmin^{-2}$) distribution for probable stars present in the decontaminated CMDs of FSR 784 (top left), Sh2-235 Cluster (top right), Sh2-235E2 (bottom left) and Sh2-235B (bottom right) respectively.}
\label{fig:10}
\end{figure*}

Fig.~\ref{fig:10} shows the number-density distribution for probable stars in the decontaminated $J\times(J-K_s)$ CMDs. The densest parts of those Hess diagrams correspond to the PMS.

The present sample of ECs suggests some age spread implied mainly by the PMS stars. This suggests some sequential star formation (Fig.~\ref{fig:11}). However, the visual extinction of the same ECs are much smaller than the values observed for individual stars in these clusters which suggests that these stars are probably embedded in dense structures, which explains the larger values of $A_V$ estimated in the previous works, especially in the case of individual star analyses.

Figs. \ref{fig:02}, \ref{fig:03} and \ref{fig:11}, together with the age derived for the Sh2-235 Cluster ($\approx5$ Myr), indicate that the O star ($\approx1$ Myr) is not a member of this cluster. A possible explanation for the presence of the O star near this cluster is that winds from Sh2-235 Cluster, colliding with the surrounding gas, might have originated this star, CBB 2 and other clusters in the neighborhood.
Assuming $v\approx20\,km\,s^{-1}$ for the dense gas involved in the expansion \citep{Kirsanova08}, the Sh2-235 Cluster might be responsible for sequential star formation across a region of radius $\approx10$ pc (Fig.~\ref{fig:11}), which includes also the clusters in a row in the southwest direction. 
Sequential star formation is possible also for the pairs G173 and CBB 1, and Sh2-233SE Cluster and PCS 2.
Recently, \citet{Dewangan11} point out that star formation continues to occur in the Sh2-235 complex, mainly within the ECs. They identified 86 Class 0/I and 144 Class II YSOs, which reinforce the possibility of a sequential star formation event.
 
\subsection{Colour-colour diagrams}
\label{sec:3.3}

 Colour-colour diagrams are useful tools to investigate the nature of ECs. We show in Fig. \ref{fig:09} the decontaminated near-IR colour-colour diagram $(J-K_s)\times(H-K_s)$ of the member stars, together with PMS tracks \citep{Siess00}, set with the reddening values derived above, to estimate ages. As a consequence of the presence of the PMS stars in the cluster, it is expected that some stars present near-IR excess. As expected from the CMDs of ECs (Figs.~\ref{fig:04} to \ref{fig:08}), a significant fraction of the stars appears to be very reddened. Most stars, specially MS stars, have $(H-K_s)$ colours close to the isochrone, within the uncertainties.  Besides, most of the very red PMS stars are displaced parallel to the respective reddening vectors. However, a significant fraction appears to present an abnormal excess in $(J-K_s)$ and $(H-K_s)$, especially Sh2-235B, which may come from PMS stars still bearing circumstellar discs. MS stars lie on the blue side of the diagrams and there occurs a gap between MS and PMS stars in the CMDs. 

\section{Cluster structure}
\label{sec:4}

The structure of the ECs is analysed by means of the stellar radial density profile (RDP), defined as the projected number of stars per area surrounding the cluster centre.
RDPs are built with stars selected after applying the respective colour magnitude (CM) filter to the observed photometry. CM filters isolate the probable cluster sequences excluding stars with colours different from those of the cluster sequences \citep[e.g.][and references therein]{Bonatto07a}. However, residual field stars with colours similar to those of the cluster are expected to remain inside the CM filter. They affect the intrinsic stellar radial distribution profile in a degree that depends on the relative densities of field and cluster. The contribution of these residual field stars to the RDPs is statistically quantified by means of comparison to the field. In practical terms, the use of the CM filters in cluster sequences enhances the contrast of the RDP with respect to the stellar field. The CM filters are shown in Figs.~\ref{fig:04} to \ref{fig:08} as the shaded area superimposed on the field-star decontaminated CMDs.

\begin{figure}
\resizebox{\hsize}{!}{\includegraphics{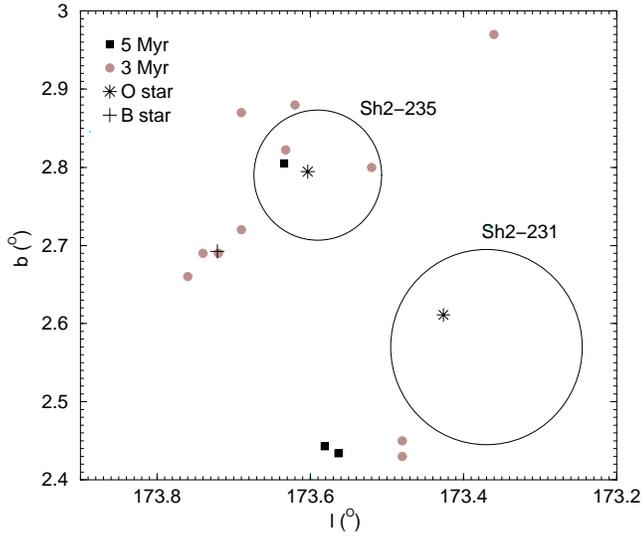}}
\caption[]{Schematic distribution of cluster positions and ages.  Asterisks are O stars, and the plus sign is a B star. It also shows the nebulae Sh2-235 and Sh2-231 as reference.}
\label{fig:11}
\end{figure}

\begin{figure}
\resizebox{\hsize}{!}{\includegraphics{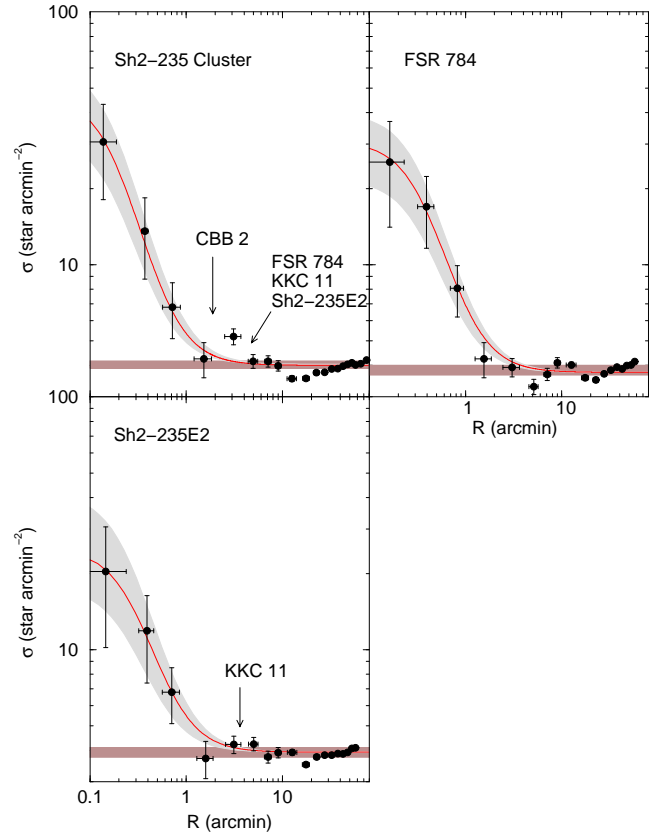}}
\caption[]{Stellar radial densities profile (filled circles) built with colour-magnitude filtered photometry, centred in the coordinates of Sh2-235 Cluster, FSR 784 and Sh2-235E2. Solid line: best-fit King profile. Horizontal shaded region: stellar background level measured in the comparison field. Gray regions: $1\sigma$ King fit uncertainty.}
\label{fig:12}
\end{figure}
 
\begin{figure}
\resizebox{\hsize}{!}{\includegraphics{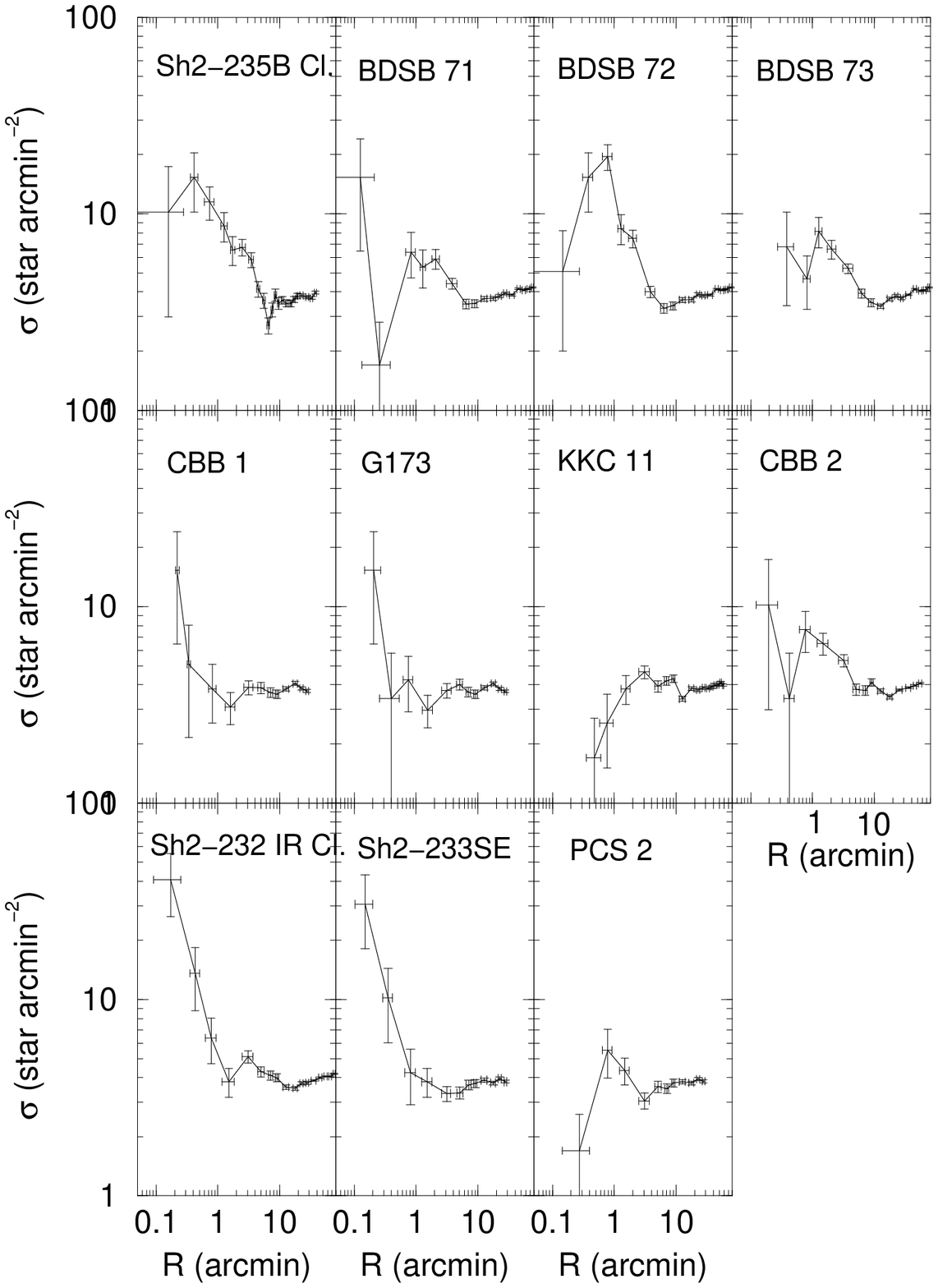}}
\caption[]{Stellar RDPs for the remaining ECs built with colour-magnitude filtered photometry.}
\label{fig:13}
\end{figure}

To minimise oversampling near the centre and under-sampling for large radii, the RDPs are built by counting stars in concentric rings of increasing width with distance to the centre. The selected number and width of rings produce RDPs with adequate spatial resolution and moderate $1\sigma$ Poisson errors. The residual background level of each RDP corresponds to the average number of CM-filtered stars measured in the comparison field. 

The cluster structure was derived by means of a  King-like profile, which is similar to a two-parameter \citet{King1962} model that describes the intermediate and central regions of globular clusters. 
The fit was performed with a non-linear least-squares routine that uses the errors as weights. The best-fit solutions are shown in Fig. ~\ref{fig:12} for 3 relatively populous ECs as a solid line superimposed on the RDPs. The profile is expressed as $\sigma(R)=\sigma_{bg}+\sigma_{0K}/(1+(R/R_{core})^{2}$, where $\sigma_{bg}$ is the stellar background surface density, $\sigma_{0K}$ is the central density relative to the background level and $R_{core}$ is the core radius. The cluster radius ($R_{RDP}$) and uncertainty can be estimated by considering the RDP fluctuations with respect to the residual field. $R_{RDP}$ is the distance from the cluster centre where RDP and comparison field become statistically indistinguishable. Small variations in the RDPs are probably due to the presence of other clusters and/or enhanced dust absorption. The derived structural parameters are given in Table~\ref{tab5}. 

In Fig.~\ref{fig:12} we show the RDPs of Sh2-235 Cluster, Sh2-235 East2, and FSR 784. For Sh2-235 Cluster and Sh2-235 East2. Overdensities show up in their RDPs. 

Fig.~\ref{fig:13} shows the RDPs of Sh2-235B Cluster, BDSB 71, BDSB 72, BDSB 73, Sh2-232 IR Cluster, KKC11, CBB 2 and the pairs CBB 1 and G 173, PCS 2 and Sh2-233 SE. The RDPs are typical of ECs of low mass and/or initial evolutionary phases, and cannot be fitted by King's profile \citep{Soares05}. They present bumps and dips as compared to field stars. The depression in star counts in the central region of some ECs is possibly due to strong dust absorption, crowding or structured cores. 
Sh2-232 IR Cluster has a high cluster/background density contrast, but the profile is irregular. The Sh2-235B cluster profile includes several of the small clusters nearby like BDSB 71, 72 and 73. A deeper photometry is required for the analysis of the structure of these objects.

\begin{table*}
{\footnotesize
\begin{center}
\caption{Structural parameters.}
\renewcommand{\tabcolsep}{3.7mm}
\renewcommand{\arraystretch}{1.3}
\begin{tabular}{lrrrrrrrr}
\hline
\hline
Cluster&$(1')$&$\sigma_{0K}$&$R_{core}$&$R_{RDP}$&$\sigma_{0K}$&$R_{core}$&$R_{RDP}$&${\Delta}R$\\
&($pc$)&($*\,pc^{-2}$)&($pc$)&($pc$)&($*\,\arcmin^{-2}$)&($\arcmin$)&($\arcmin$)&($\arcmin$)\\
($1$)&($2$)&($3$)&($4$)&($5$)&($6$)&($7$)&($8$)&($9$)\\
\hline
FSR\,784 &$0.69$&$56.7\pm4.0$&$0.25\pm0.01$&$2.1\pm0.7$&$27.02\pm1.9$&$0.36\pm0.02$&$3.0\pm1.0$&$20-60$\\
Sh2-235\,E2 &$0.60$&$71.7\pm25.0$&$0.13\pm0.03$&$1.2\pm0.3$&$25.8\pm9.0$&$0.21\pm0.05$&$2.0\pm0.5$&$20-80$\\
Sh2-235\,Cl. &$0.56$&$138.1\pm27.4$&$0.10\pm0.01$&$1.4\pm0.3$&$43.32\pm8.6$&$0.18\pm0.02$&$2.5\pm0.5$&$20-80$\\
\hline
\end{tabular}
\begin{list}{Table Notes.}
\item Col. 2: arcmin to parsec scale. To minimise degrees of freedom in RDP fits with the King-like profile (see text), $\sigma_{bg}$ was kept fixed (measured in the respective comparison fields) while $\sigma_{0}$ and $R_{core}$ were allowed to vary. Col. 9: comparison field ring. 
\end{list}
\label{tab5}
\end{center}
}
\end{table*}

\begin{table*}
\caption[]{Stellar mass estimate for more populous ECs}
\label{tab8}
\renewcommand{\tabcolsep}{3.8mm}
\renewcommand{\arraystretch}{1.25}
\begin{tabular}{cccccccccccc}
\hline\hline
&&\multicolumn{4}{c}{MS}&&\multicolumn{2}{c}{PMS}&&\multicolumn{2}{c}{$MS+PMS$}\\
\cline{3-6}\cline{8-9}\cline{11-12}
Cluster&&$\Delta\,m_{MS}$&&$N$&$M$   &&$N$&$M$ &&$N$&$M$\\
     && ($M_\odot$)    &      &(stars) &($M_\odot$)&&(stars)  &($M_\odot$)&&(stars) &($M_\odot$)     \\
(1)&&(2)&&(3)&(4)&&(5)&(6)&&(7)&(8)\\
\hline
FSR\,784&&2.90-4.30&&$3\pm1$&$9\pm4$&&$65\pm15$&$39\pm9$&&$68\pm16$&$48\pm13$\\
Sh2-235E2&&2.50-5.25&&$6\pm2$&$24\pm9$&&$41\pm6$&$25\pm4$&&$47\pm8$&$49\pm13$\\
Sh2-235\,Cl.&&1.30-6.25&&$5\pm2$&$22\pm7$&&$67\pm15$&$40\pm9$&&$72\pm17$&$62\pm16$\\
\hline
\end{tabular}
\begin{list}{Table Notes.}
\item Col. 2: MS mass range. Cols. 3-6: stellar content of the MS and PMS stars. Cols. 7-8: total (MS+PMS) stellar content.
\end{list}
\end{table*}

\begin{table*}
{\footnotesize
\begin{center}
\caption{Integrated colours and magnitudes.}
\label{tab6}
\renewcommand{\tabcolsep}{4.3mm}
\renewcommand{\arraystretch}{1.3}
\begin{tabular}{lrrrrrrrrrr}
\hline
\hline
\multicolumn{8}{c}{Magnitude}&\multicolumn{1}{c}{}&\multicolumn{2}{c}{Colour}\\
\cline{2-8}
\cline{10-11}
\multicolumn{1}{c}{}&\multicolumn{3}{c}{Apparent}&\multicolumn{1}{c}{}&\multicolumn{3}{c}{Absolute}&\multicolumn{1}{c}{}&\multicolumn{2}{c}{Reddening Corrected}\\
\cline{2-4}
\cline{6-8}
\cline{10-11}
Cluster&$J$&$H$&$K_s$&&$J$&$H$&$K_s$&&$(J-H)$&$(J-K_s)$\\
(1)&(2)&(3)&(4)&&(5)&(6)&(7)&&(8)&(9)\\
\hline
FSR\,784 &12.9&9.7&8.8&&-0.02&-2.8&-4.0&&$2.82\pm0.14$&$3.57\pm0.14$\\
Sh2-235\,E2 &13.2&11.7&10.6&&0.5&-0.6&-1.5&&$1.11\pm0.33$&$1.99\pm0.31$\\
Sh2-235\,Cl.&8.5&8.2&7.9&&-4.0&-4.0&-4.0&&$-0.05\pm0.04$&$0.01\pm0.04$\\ 
\hline
\end{tabular}
\begin{list}{Table Notes.}
\item Col. 2-4: apparent magnitude. Cols. 5-7: absolute magnitude. Cols. 8-9: $(J-H)$ and $(J-K_s)$  colours.  
\end{list}
\end{center}
}
\end{table*}

\begin{figure}
\begin{minipage}[b]{1.0\linewidth}
\includegraphics[width=\textwidth]{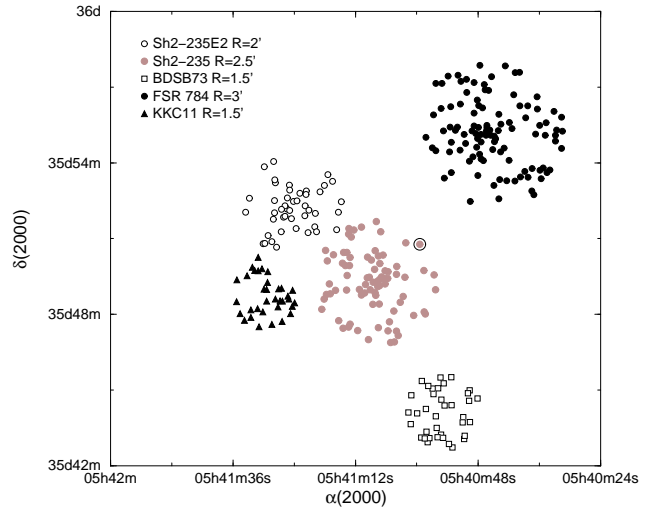}
\end{minipage}\hfill
\caption[]{Angular distribution of the decontaminated stellar content for FSR 784, Sh2-235, Sh2-235E2, KKC 11 and BDSB 73. BD$+35^{\circ}1201$ is indicated as an open circle around the star.}
\label{fig:14}
\end{figure}

The structure of young populous ECs can be generally characterised by an RDP with multiple peaks on a large spatial scale or centrally condensed with an RDP that can be described by a King's law \citep{Lada03}, although the present objects are not virialised. However, the differential dust absorption produces conspicuous variations in the RDPs of the present objects.
King profile describes the structure of clusters close to spherical symmetry and centrally concentrated. However, many young clusters are substructured or asymmetric, deviating significantly from this shape, and therefore cannot be fitted by King's law \citep{Cartwright04, Gutermuth05}. In Fig.~\ref{fig:14}, we show the spatial distribution of stars in the decontaminated photometry of the three ECs that follow a King-like profile (FSR 784, Sh2-235 Cluster and Sh2-235E2) and two representative cases of objects that do not (KKC 11 and BDSB 73). The former are centrally concentrated and nearly circularly symmetric. The cavities and overdensities in the stellar distribution (Fig.~\ref{fig:14}) can be seen as bumps and dips in the RDP (Fig.~\ref{fig:11}). On the other hand, objects like BDSB 73 that are not centrally condensed and KKC 11 with more elongated shape do not follow a King profile. The multiple peaks in the RDPs of these ECs may be a fractal effect. If these objects survive the primordial gas expulsion, they may undergo merging evolving into a relatively smooth structure.

The angular distribution of decontaminated stars (Fig.~\ref{fig:14}), used in the CMD construction, reproduces the distribution of stars in the RDPs built with filtered photometry (Figs.~\ref{fig:12} and \ref{fig:13}), supporting the consistency of our results.

\section{Mass estimates}
\label{Mass}
Given the poorly-populated nature of the MS, we simply counted stars in the CMDs (within the region $R<R_{RDP}$), and summed their masses as estimated from the mass-luminosity relation implied by the respective isochrone solution (Sect. \ref{sec:3.2}). The results are given in Table \ref{tab8}.

All the ECs clearly present distinct populations of MS and PMS stars (Figs.~\ref{fig:04} to \ref{fig:08}).
However, given the differential reddening, it is not possible 
to attribute a precise mass value for each PMS star. Thus, we simply count the number of PMS 
stars and adopt an average mass value for the PMS stars to estimate $n_{PMS}$ and $m_{PMS}$.
Assuming that the mass distribution of the PMS stars also follows Kroupa's (2001) MF, the zsc,
average PMS mass - for masses within the range $0.08\la m(\ms)\la7$ - is $<m_{PMS}>\approx0.6\ms$.
Thus, we simply multiply the number of PMS stars (Table~8) by this value to estimate the PMS 
mass. Finally, we add the latter value to the MS mass to obtain an estimate of the total 
stellar mass. These values should be taken as lower limits. 

\section{Relations among astrophysical parameters}
\label{sec:5}

\begin{figure}
\resizebox{\hsize}{!}{\includegraphics{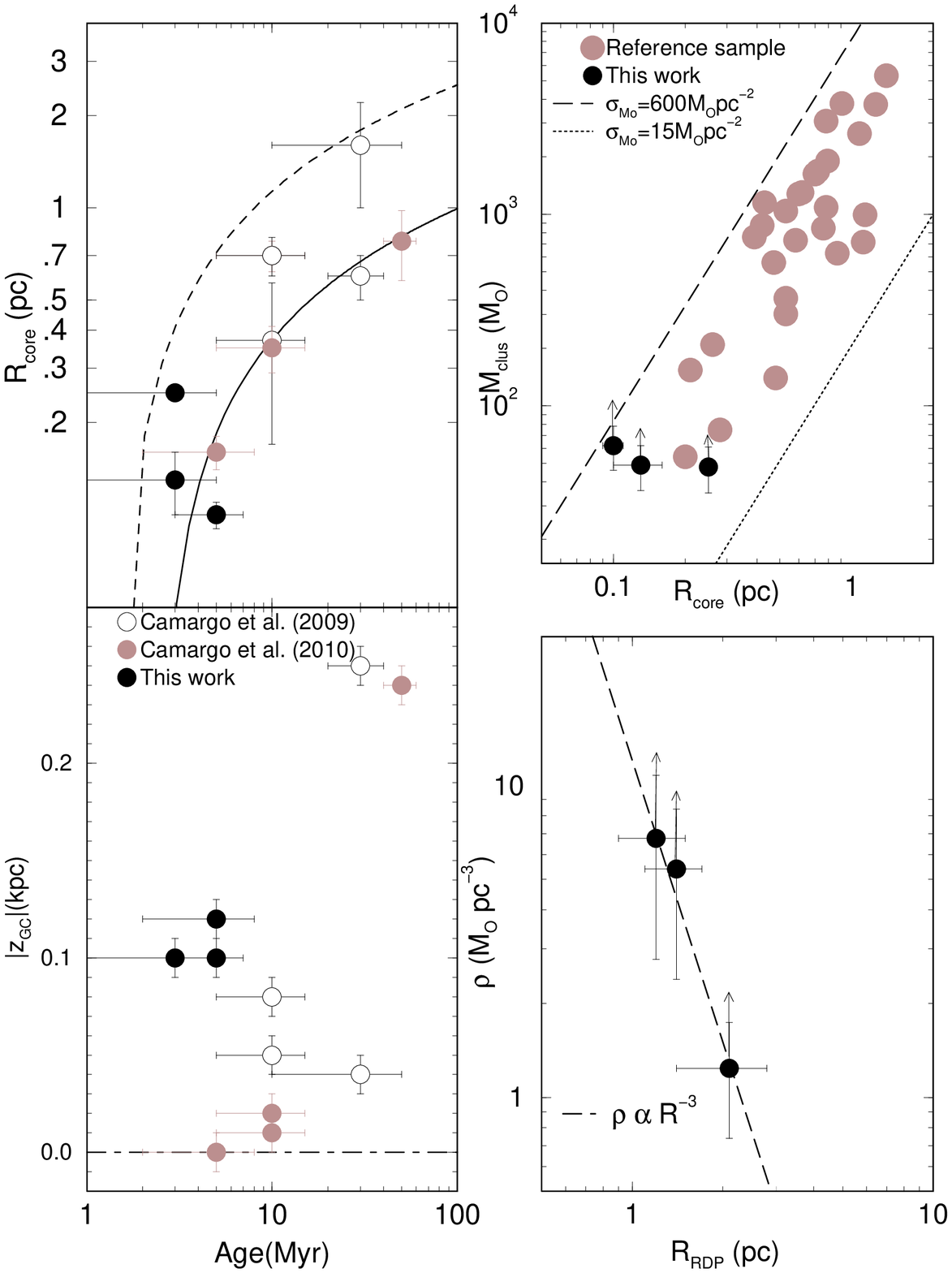}}
\caption[]{Present cluster properties compared to literature ones. Top-left panel: age and $R_{core}$ relations. The dashed line is the logarithmic fit of \citet{Bastian08} 
($R_{core}(pc)=0.6\times{ln}(age[Myr])-0.25$) for M51 clusters. The solid line is our fit for Galactic clusters. 
Bottom-left panel: $|z_{GC}|$ for clusters of the top panel. Top right: core radius and cluster mass follow the 
relation $M_{clus}=13.8\sigma_{M0}R^{2}_{c}$, with varying values of $\sigma_{M0}$; reference sample OCs are shown
as gray circles. Bottom right: cluster density {\em vs.} radius. Arrows indicate lower limits to the cluster mass and density.}
\label{fig:15}
\end{figure}

N-body simulations of massive star clusters that include the effect of gas removal \citep[e.g.][]{Goodwin06} show that the phase of dramatic core radii increase may last about 10-30 Myr. Mass segregation may also lead to a phase of core contraction, with high mass stars more concentrated in the core while low mass stars are transferred to outer parts of the cluster. In this context, we suggest that most objects of our sample have not yet removed completely their primordial gas and are not in this expansion phase. As a consequence the core and cluster radii remain small. However, these ECs may be intrinsically small as well.

\citet{Lada03} argue that the structure in ECs reflects the underlying structure in the dense molecular gas from which they formed and suggest two possible structures for these objects: those similar to classical OCs (radially concentrated that fit a King law) and those that exhibit a density profile with multiple peaks. The latter kind of ECs present a fractal-like structure with smaller substructures or mini clusters that are probably a consequence of the fact that these objects are formed in a GMC with a fractal substructure  \citep[][and references therein]{Schmeja08, Lomax10, Sanchez10}. They evolve with time and their fates depend on the processes of dissolution, which they will undergo.  If the ECs do not endure the action of the dissolution processes, they might evolve into a homogeneous distribution of stars and eventually disperse. Otherwise their fate will be a centrally concentrated distribution of stars or a bound cluster. This is possible for the lower mass ECs and cluster pairs or multiples (Figs.~\ref{fig:04} to \ref{fig:08}).  
It would also be important to have deeper observations  of  the less-populated clusters with irregular RDPs  (Fig.~\ref{fig:13}). Probing fainter PMS stars
might contribute to the construction of better sampled stellar density profiles, like those of  the intrinsically populous and  massive clusters in the present star-forming complex (Fig.~\ref{fig:12}).

When the projected mass density of a star cluster 
follows a King-like profile \citep{StrucPar}, the cluster mass ($\rm M_{clus}$) 
can be expressed as a function of the core radius and the central surface mass-density
($\sigma_{M0}$) according to $\rm M_{clus}\approx13.8\sigma_{M0}\,R^2_{C}$ \citep{Bonatto09}. Fig.~\ref{fig:15}
(top-right panel) shows the distribution of our populous ECs in the plane core radius (Sect.~\ref{sec:4}) 
{\em vs} cluster mass (Sect.~\ref{Mass}). Clearly, our ECs (together with the reference sample)
distribute parallel to the above relation, being constrained within King-like distributions 
with $\rm15\la\sigma_{M0}\,(\ms\,pc^{-2})\la600$ (these limits take the uncertainties into 
account). This correlation between cluster mass and core radius is consistent 
with the mass-radius relation suggested by \citet{PZ10} for massive clusters younger than 
100\,Myr and extended for less massive ones \citep{Camargo10}. 

Since, the cluster radius and mass were estimated, we compute the cluster mass density $\rm\rho(\ms\,pc^{-3})=\frac{3}{4\pi}M_{clus}\,R^{-3}_{RDP}$. The results are shown in the plane $R_{RDP}\,vs.\,\rho$ (Fig.~\ref{fig:15}, bottom-right panel). Despite the error bars, the density decreases with cluster radius as $\rho\propto R^{-(3.0\pm0.3)}_{RDP}$, similarly to the cluster sample studied by \citet{Camargo10}. We also show in Fig.~\ref{fig:15} the relation between age and both $R_{core}$ and $|Z_{GC}|$ for the present ECs. We fit an empirical curve $R_c(pc)=0.27\times{ln}\,(age[Myr])-0.25$ (the solid line) to Galactic OCs younger than 100 Myr. The dashed line is the observed relation for M51 clusters \citep{Bastian08}. The present clusters behave as relatively low mass ECs, as expected. 

\section{Concluding remarks}
\label{sec:6}

In the present work we performed a field-star decontaminated 2MASS analysis of 14 ECs in the H II regions  Sh2-235, Sh2-233, Sh2-232 and Sh2-231, and other small nebulae in the area.
We were able to derive astrophysical parameters and investigate the nature of young clusters embedded in these H II regions. Fundamental parameters were obtained for all ECs analysed, but structural parameters were derived for FSR 784, Sh2-235E2 and Sh2-235 Cluster. The decontaminated CMDs exhibit a poorly-populated MS and a large fraction of PMS stars, and suggest some age spread, as expected for sequential star formation process. Two new ECs (CBB 1 and CBB 2) were discovered in this work.

The age, size and location of KKC 11, FSR 784, Sh2-235 Cluster, CBB 2, and Sh2-235E2 are consistent with the \textit{collect and collapse} scenario. 
The CMDs of Sh2-235 B, BDSB 71, BDSB 72, BDSB 73, Sh2-232 IR Cluster, G173, CBB 1, and PCS 2 suggest that these objects are ECs. The enhanced dust absorption mainly in the innermost region indicates that deep photometry is necessary to derive their structural parameters. 

The present ECs have core and cluster radii smaller than clusters at the same Galactocentric distance and age. Probably most of them have not expelled the primordial gas completely, and thus have not expanded. 

\vspace{0.8cm}

\textit{Acknowledgements}: We thank an anonymous referee for important comments and suggestions. This publication makes use of data products from the Two Micron All Sky Survey, which is a joint project of the University of Massachusetts and the Infrared Processing and Analysis Centre/California Institute of Technology, funded by the National Aeronautics and Space Administration and the National Science Foundation. This research has made use of the WEBDA database, operated at the Institute for Astronomy of the University of Vienna, as well as Digitised Sky Survey images from the Space Telescope Science Institute obtained using the extraction tool from CADC (Canada). We acknowledge support from CNPq and CAPES (Brazil).

\label{lastpage}
\end{document}